\documentclass{pasj00}
\draft

\begin{document}
\SetRunningHead{Takakuwa \& Kamazaki}{Submillimeter Lines in Protostellar Envelopes}
\Received{2010/11/18}%{yyyy/mm/dd}
\Accepted{2011/05/08}%{yyyy/mm/dd}
%\Published{}%{yyyy/mm/dd}

\title{Skewed Distributions and Opposite Velocity Gradients of Submillimeter Molecular Lines\\
in Low-Mass Protostellar Envelopes}

%%% begin:list of authors
% Do NOT capitalize all letters in "textsc".
\author{Shigehisa \textsc{Takakuwa}} %
%  \thanks{Example: Present Address is xxxxxxxxxx}}
\affil{Academia Sinica Institute of Astronomy and Astrophysics, P.O. Box 23-141, Taipei 10617, Taiwan}
\email{takakuwa@asiaa.sinica.edu.tw}

\and

\author{Takeshi \textsc{Kamazaki}}
\affil{Joint ALMA Observatory, Av. Alonso de Cordova 3107, Vitacura, Santiago, Chile}
\email{tkamazaki@alma.cl}

%%% Please use the following style in case that sorting by 
%%% affiliation is impossible. 
%
% \author{%
%   D-Firstname \textsc{D-Familyname}\altaffilmark{1}
%   E-Firstname \textsc{E-Familyname}\altaffilmark{1,2}
%   and
%   F-Firstname \textsc{F-Familyname}\altaffilmark{2}}
% \altaffiltext{1}{Address of Institute}
% \email{ddddd@xxx.xxx.xx.xx}
% \email{eeeee@xxx.xxx.xx.xx}
% \altaffiltext{2}{Address of Institute}

%% `\KeyWords{}' always has to be placed before `\maketitle'.
%Do NOT move this preamble from here!

\maketitle

\begin{abstract}

We have made mapping observations of L1551 IRS 5, L1551NE, L723, and L43
and single-point observations of IRAS 16293-2422 in the submillimeter CS ($J$ = 7--6)
and HCN ($J$ = 4--3) lines with ASTE.
Including our previous ASTE observations of L483 and B335,
% we detected the submillimeter lines above 4$\sigma$ level (1$\sigma$ = 0.1 - 0.2 K)
% toward all the protostellar positions except for L1551 NE and L43,
% and
we found a clear linear correlation between the source bolometric luminosities and the total integrated intensities of
the submillimeter lines ($I_{CS}$ $\propto$ $L_{bol}^{0.92}$).
The combined ASTE + SMA CS (7--6) image of L1551 IRS 5 exhibits
an extended ($\sim$2000 AU) component tracing the associated reflection nebula at the west and southwest,
as well as a compact ($\lesssim$500 AU) component centered on the protostellar position.
The emission peaks of the CS and HCN emissions in L1551 NE are not located
at the protostellar position but offset ($\sim$1400 AU) toward the associated reflection nebula at the west.
% A similar extended feature of the
% submillimeter CS and HCN emissions in L483 has already been reported in our early paper.
With the statistical analyses,
we confirmed the opposite velocity gradients of the CS (7--6) emission
to those of the millimeter lines along the outflow direction,
which we reported in our early paper. The magnitudes of the submillimeter velocity gradients
are estimated to be (9.7$\pm$1.7) $\times$ 10$^{-3}$ km s$^{-1}$ arcsec$^{-1}$
in L1551 IRS 5 and (7.6$\pm$2.4) $\times$ 10$^{-3}$ km s$^{-1}$ arcsec$^{-1}$ in L483.
% With the statistical analyses,
% we confirmed the opposite velocity gradients of the CS (7--6) emission
% to those of the millimeter lines or the associated molecular outflows
% along the outflow direction, 
% in L1551 IRS 5 ((9.7$\pm$1.7) $\times$ 10$^{-3}$ km s$^{-1}$ arcsec$^{-1}$) and
% L483 ((7.6$\pm$2.4) $\times$ 10$^{-3}$ km s$^{-1}$ arcsec$^{-1}$),
% which we reported in our early paper.
% we verified the opposite velocity gradients
% of the CS (7--6) emission along the outflow direction
% to those of the millimeter lines or the associated molecular outflows,
% which we reported in our early paper, in L1551 IRS 5 ((9.7$\pm$1.7) $\times$ 10$^{-3}$ km s$^{-1}$ arcsec$^{-1}$) and
% L483 ((7.6$\pm$2.4) $\times$ 10$^{-3}$ km s$^{-1}$ arcsec$^{-1}$).
We suggest that the ``skewed'' submillimeter molecular emissions toward the associated reflection nebulae
at a few thousands AU scale trace the warm ($\gtrsim$40 K) walls of the envelope cavities,
excavated by the associated outflows and irradiated by the central protostars directly.
The opposite velocity gradients along the outflow direction likely
reflect the dispersing gas motion at the wall of the cavity in the envelopes perpendicular to the outflow.
% The detected linear correlation between the protostellar luminosities and the intensities
% of the submillimeter molecular emissions at a few thousand AU scale may support this
% interpretation.

\end{abstract}

\section{Introduction}

Low-mass stars form in 3000 - 10000 AU scale condensations of molecular gas and dusts, so-called dense cores \citep{and00,mye00}.
Previous interferometric observations in millimeter molecular lines have revealed rotating and infalling gas motions
in dense cores associated with protostars, ``protostellar envelopes" \citep{oh96b,oh97a,oh97b,mom98}.
These millimeter observations have probed structures and kinematics of
molecular gas with a temperature of $\sim$10 - 20 K and a density of 10$^{4-6}$ cm$^{-3}$ in the envelopes.
On the other hand, submillimeter molecular lines can trace warmer ($\gtrsim$40 K) and denser
($\gtrsim$10$^{7}$ cm$^{-3}$) regions, and
recent interferometric observations of protostellar envelopes in submillimeter molecular lines with the SMA
have revealed compact ($\lesssim$500 AU) components associated with the central protostars,
which often show rotational (possibly Keplerian) gas motion around the protostars
\citep{tak04,ta07b,bri07,lom08,jor09,yen11}.
% Combining these millimeter and submillimeter study of protostellar envelopes is a key to
% understand how envelope gas over a wide range of physical conditions accretes and transfers angular momentum
% to form the star and disk system at the center.

The SMA observations of protostellar envelopes, however, suffer from the effect of the missing flux.
For example, the SMA observations of L1551 IRS 5 in the CS ($J$ = 7--6) line \citep{tak04} and those of
IRAS 16293-2422 in the HCN ($J$ = 4--3) line
\citep{ta07b} recovered only $\sim$11 $\%$ and $\sim$35 $\%$ of the total fluxes observed with CSO and JCMT,
respectively, suggesting the presence of the extended ($\gtrsim$2000 AU) submillimeter components.
In fact, the combined SMA + JCMT image of IRAS 16293-2422 in the HCN (4--3) line
shows an extended ($\sim$3000 AU) envelope structure as well as a compact ($\sim$500 AU) disklike structure
associated with the protostar.
These results suggest that there is extended ($\gtrsim$2000 AU), warm ($\gtrsim$40K) and/or
dense ($\gtrsim$10$^{7}$ cm$^{-3}$) gas in protostellar envelopes.

In order to investigate the origin of the extended submillimeter molecular emissions in protostellar envelopes,
we initiated mapping observations of protostellar envelopes in the CS (7--6) and HCN (4--3) lines
with Atacama Submillimeter Telescope Experiment
(ASTE)\footnote{The ASTE project is driven by Nobeyama Radio Observatory (NRO), a branch
of National Astronomical Observatory of Japan (NAOJ), in collaboration
with University of Chile, and Japanese institutes including University
of Tokyo, Nagoya University, Osaka Prefecture University, Ibaraki
University., and Hokkaido University.}, a 10-m submillimeter single-dish telescope at the Atacama Site in Chile (\cite{ta07a}; hereafter Paper I).
In Paper I, we mapped the protostellar envelopes around L483 and B335.
In L483, the HCN emission is slightly resolved and exhibits a western extension ($\sim$5000 AU)
toward the direction of the associated reflection nebula and the blueshifted outflow.
Furthermore, the position-velocity diagrams of the HCN and CS emissions
along the axis of the associated molecular outflows in L483 and B335
exhibit possible velocity gradients opposite to those
of the millimeter emissions or the associated molecular outflows.
% These results imply that we need to reconsider the origin of the submillimeter molecular emissions
% in low-mass protostellar envelopes.

These results in Paper I are, however, still a little marginal and more robust confirmations of those results,
including more observing samples and more sophisticated data analyses, are required.
In the present paper, we have expanded our previous ASTE observations and have performed
mapping observations of L1551 IRS 5, L1551 NE, L723, and L43.
All of these and previous source samples are nearby ($D$ $\lesssim$300 pc) and
representative ($L_{bol}$ $\gtrsim$3 $\LO$)
low-mass protostellar objects suitable for the purpose of the present study.
We have constructed
data analysis tools by ourselves and have performed more objective, unambiguous data
analyses of the previous and the present new data.
From these observations and data analyses, we have revealed that the submillimeter molecular lines
often show ``skewed'' emission distributions tracing the reflection nebula (L1551 IRS 5, L483, L1551 NE,
and L723), and have found a linear correlation between the intensity of the submillimeter molecular lines 
and the protostellar luminosities.
We have also verified the presence of the opposite velocity gradients of the submillimeter
CS line to those of the millimeter lines and the associated outflows in L1551 IRS 5 and L483.
With these results obtained from our ASTE observations and
data analyses, we discuss the origin of the submillimeter molecular emissions in low-mass protostellar envelopes.

The structure of the present paper is as follows. In $\S$2, we describe our new ASTE observations,
and combination of the ASTE CS (7--6) data with the previously-published SMA CS (7--6) data in L1551 IRS 5
\citep{tak04}.
In $\S$3.1., we discuss all the observed spectra toward the protostellar positions,
while in $\S$3.2. we show the new results of the mapping observations.
In $\S$4.1., we discuss the extents and distributions of the submillimeter molecular lines
and their origin, including our simple radiative-transfer model.
In $\S$4.2., we verify the presence of the opposite velocity gradients with our statistical analyses, 
and discuss a possible reason for the opposite submillimeter velocity gradients.

\section{ASTE Observations and Combining with the SMA Data}

With the ASTE 10-m telescope we made mapping observations of L1551 IRS 5, L1551 NE, L723, and L43,
in the CS ($J$ = 7--6; 342.882866 GHz) and HCN ($J$ = 4--3; 354.505480 GHz) lines
on 2006 August 26, 28 -- September 2.
Remote observations were performed from an ASTE operation room of NAOJ at Mitaka, Japan, using the network
observation system N-COSMOS3 developed by NAOJ (\cite{kam05}). Details of the ASTE telescope are presented by Ezawa et al. (2004).
% Table \ref{tab:obs} summarizes the observational parameters.
A cartridge-type 350 GHz receiver mounted on ASTE is a double sideband instrument with an IF frequency range 
of 5 to 7 GHz (\cite{koh05}), and the CS and HCN lines were observed simultaneously at different sidebands.
% The telescope beam size, main beam efficiency, and the spectral resolution were $\sim$22$\arcsec$,
% $\sim$0.6, and 125 kHz ($\sim$0.11 km s$^{-1}$), respectively.
The telescope beam size and the spectral resolution were $\sim$22$\arcsec$ and
125 kHz ($\sim$0.11 km s$^{-1}$), respectively.
The typical DSB system noise temperature was $\sim$270 - 500 K.
The telescope pointing was checked at the beginning and the middle of the observations by making
continuum observations of Uranus or Jupiter, or five-point CO ($J$ = 3--2) observations of a late-type star, O-Cet,
and was found to be better than 2$\arcsec$ during the whole observing period.

As a standard source we also observed the central positions of IRAS 16293-2422 and L1551 IRS 5,
and confirmed that the relative intensity was consistent within $\sim$30$\%$.
The total on-source integration times for IRAS 16293-2422 and L1551 IRS 5 were 11.8 minutes
and 23.8 minutes, respectively.
We compared the observed CS and HCN spectra toward IRAS 16293-2422
to those obtained with the CSO telescope which has the same dish size as that of the ASTE
telescope (\cite{bla94,dis95}), and found that the main beam efficiency of ASTE was
$\sim$0.6 both for the CS and HCN lines. No further correction for the sideband ratio was performed.
Hereafter we show the observed line intensities in the unit of $T_{MB}$.

In the mapping observations, typical rms noise levels per point were $\sim$0.16 K, $\sim$0.20 K, $\sim$0.20 K,
and $\sim$0.27 K in L1551 IRS 5, L1551 NE, L723, and L43, respectively, with a typical on-source integration time
of $\sim$4 - 12 minutes. The mapping regions cover most of the CS and HCN emission regions at a grid
spacing of 10$\arcsec$, providing the Nyquist-sampled maps. Toward the center of L43
a better rms noise level ($\sim$0.18 K) was obtained with an on-source integration time of 13.3 minutes,
to compare the central spectra to those toward other sources.
In Paper I, we performed mapping observations of
L483, B335, and single-point observations of L723 and IRAS 16293-2422, and in total we observed
seven protostellar sources in the submillimeter CS and HCN lines with ASTE. The observed source properties
and the ASTE observations from Paper I and the present paper (hereafter Paper II) are summarized in
Table \ref{tab:source}. 
Figure \ref{fig:I16spec} shows the CS and HCN spectra toward IRAS 16293-2422
after combining both the Paper I and II data, and Figure \ref{fig:spall}
shows the observed CS ($J$ = 7--6) and HCN ($J$ = 4--3) spectra toward
the central positions of all the other protostellar sources.

% We compared the observed HCN and CS spectra toward IRAS 16293-2422
% to those obtained with the CSO telescope which has the same dish size as that of the ASTE
% telescope (\cite{bla94,dis95}), and found that the main beam efficiency of
% the ASTE telescope was $\sim$0.6.
% We also confirmed that the asymmetric spectral shapes
% in IRAS 16293-2422 taken with the ASTE telescope
% are in excellent agreement with those taken with the CSO telescope.

% L1551 IRS 5   ~ 10 min
% CS rms (TMB) ~ 0.15 K   center ~ 0.099 K    (23m50sec)
% HCN rms (TMB) ~ 0.16 K center ~ 0.117 K
%
% L1551 NE  ~ 5 min
% CS rms (TMB) ~ 0.19 K      center ~ 0.188 K (5min)
% HCN rms (TMB) ~ 0.20 K   center ~ 0.199 K
%
% L723 ~ 12 min
% CS rms (TMB) ~ 0.19 K      center 0.188 K (12min40sec)
% HCN rms (TMB) ~ 0.21 K  center 0.188 K
%
% L43 ~ 3min20sec - 6min10sec
% CS rms (TMB) ~ 0.25 K        center 0.177 K (13min20sec)
% HCN rms (TMB) ~ 0.28 K     center 0.186 K ()
We also combined the ASTE CS (7--6) image of L1551 IRS 5 with the SMA CS (7--6) image \citep{tak04},
adopting the method described by \citet{ta07b}. Details of the SMA observations are described in \citet{tak04}.
The conversion factor from $T_{MB}$ (K) to S (Jy beam$^{-1}$) was derived to be 46.6 as
\begin{equation} 
S = \frac{2k_{B}\Omega_{beam}}{\lambda^2}T_{MB},
\end{equation}
where $k_{B}$ is the Boltzmann constant, $\lambda$ is the wavelength,
and $\Omega_{beam}$ is the solid angle of the ASTE beam (= 22$\arcsec$).
The SMA observations recovered $\sim$10$\%$ of the total CS (7--6) flux
observed with ASTE. The resultant synthesized beam size and
the rms noise level per channel in the combined image are 3$\farcs$6 $\times$
2$\farcs$7 (P.A. = -65$^{\circ}$) and $\sim$1.1 Jy beam$^{-1}$,
respectively, where the velocity resolution is the same as that of the SMA
data ($\sim$0.179 km s$^{-1}$).

% TMB --> Jy beam-1    1 (K) = 46.6 (Jy 22" beam-1)
% synthesized beam; natural weighting ~ 3.557 x 2.695 arcsec; -64.8 deg
%  rms ~ 1.1 Jy beam-1
% vel. reso ~ same as SMA ~ 0.179 km s-1
%
% missing flux   ---> Need to convolve SMA map with the ASTE beam
% recovered flux ~ 10 %

\begin{table}
\footnotesize
\begin{center}
\caption{Observed sources}\label{tab:source}
\begin{tabular}{llccccccl}
\hline\hline
Source &IRAS &$L_{bol}$ &$T_{bol}$ &R. A. &Dec. &Distance &Ref. \footnotemark[$*$] &ASTE Observation\\
              &           &($\LO$) &(K)       &(J2000.0)   &(J2000.0)   &(pc) & & \\
\hline
L1551 IRS 5      &04287+1801 &22  &92  &04$^{h}$31$^{m}$34$^{s}.$14 &18$^{\circ}$08$\arcmin$05$\farcs$1  &140  &1,2,3,4 &Paper II (mapping)\\
L1551 NE        &04289+1802   &4.2 &91  &04 31 44.47      &18 08 32.2   &140       &1,2,4,5 &Paper II (mapping)\\
IRAS 16293-2422 &16293-2422 &21  &42  &16 32 22.87      &-24 28 36.6  &160       &2,6,7 &Paper I \& II (both one point)\\
L43             &16316-1540 &2.7 &370 &16 34 29.30      &-15 47 01.7  &125       &8,9 &Paper II (mapping)\\
L483            &18148-0440 &13  &52  &18 17 29.86      &-04 39 38.7  &200       &2,8,10 &Paper I (mapping) \\
L723            &19156+1906 &3.3 &47  &19 17 53.62      &19 12 19.5   &300       &2,8,11 &Paper I (one point) \& II (mapping)\\
B335            &19345+0727 &3.1 &28  &19 37 00.94      &07 34 08.8   &150       &2,8,12 &Paper I (mapping)\\
  \hline
       \multicolumn{9}{@{}l@{}}{\hbox to 0pt{\parbox{85mm}{\footnotesize 
       \par\noindent
       \footnotemark[$*$] (1) Emerson et al. 1984; (2) Froebrich 2005; (3) Takakuwa et al. 2004; (4) Saito et al. 2001;
(5) Moriarty-Schieven et al. 2000; (6) Gregersen et al. 1997; (7) Takakuwa et al. 2007b;
(8) Shirley et al. 2000; (9) Chen et al. 2009; (10) Park et al. 2000;
(11) Anglada et al. 1991; (12) Yen, Takakuwa, $\&$ Ohashi 2010a
     }\hss}}
\end{tabular}
\end{center}
\end{table}

\section{Results}
\subsection{Submillimeter CS and HCN Spectra}

\begin{figure}
  \begin{center}
    \FigureFile(80mm,80mm){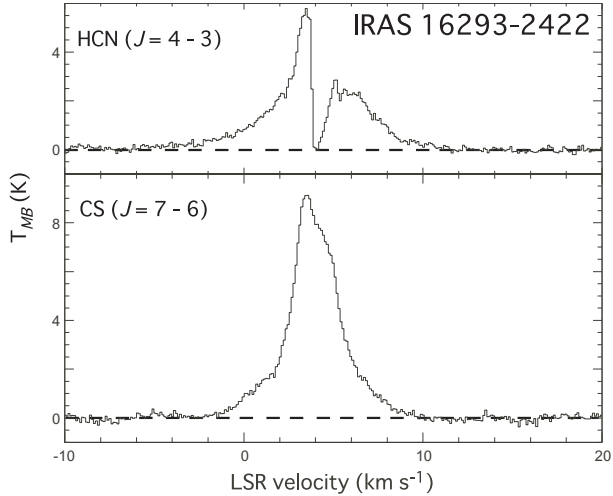}
    %%% \FigureFile(width,height){filename}
  \end{center}
  \caption{CS ($J$ = 7--6) and HCN ($J$ = 4--3) spectra toward IRAS 16293-2422
observed with ASTE.}\label{fig:I16spec}
\end{figure}

\begin{figure}
  \begin{center}
    \FigureFile(160mm,160mm){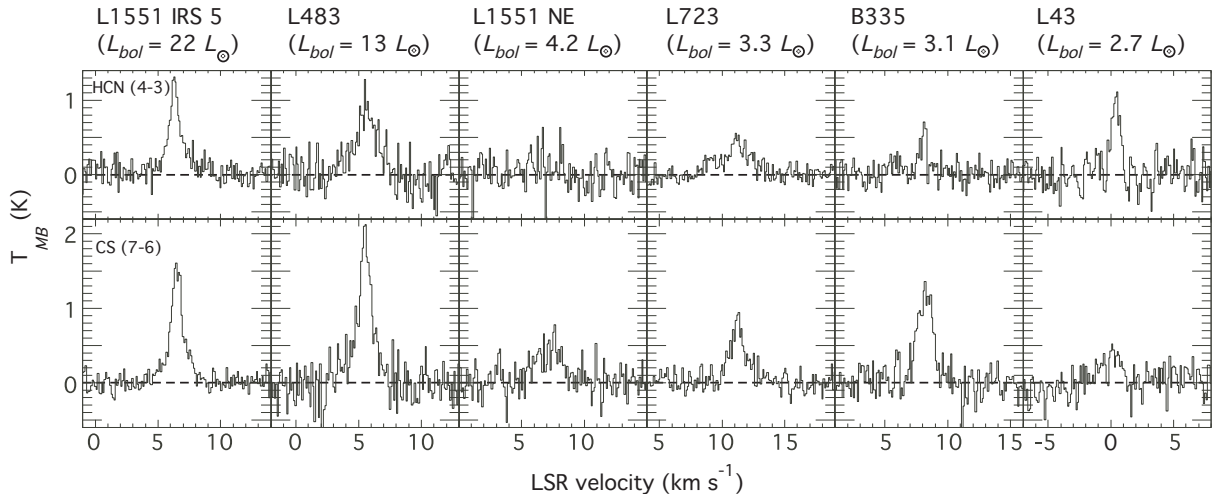}
    %%% \FigureFile(width,height){filename}
  \end{center}
  \caption{CS ($J$ = 7--6) ($lower$) and HCN ($J$ = 4--3) ($upper$) spectra
toward the protostellar positions, observed with ASTE.}\label{fig:spall}
\end{figure}

Figures \ref{fig:I16spec} and \ref{fig:spall} show the observed ASTE CS ($J$ = 7--6) and HCN ($J$ = 4--3) spectra
toward the protostellar positions of all of our targets,
and Table \ref{tab:spectra} summarizes the observed line parameters derived from Gaussian fittings
unless otherwise noted.
Except for the CS and HCN lines toward L1551 NE and the CS line toward L43,
we detected these submillimeter lines above 4$\sigma$ level.
The CS line is stronger than the HCN line except for L43.
The CS intensity ranges from $\lesssim$0.5 K (L1551 NE) to $\sim$9.1 K (IRAS 16293-2422)
and the HCN intensity from
$\sim$0.4 K (L723) to $\sim$5.8 K (IRAS 16293-2422). % except for those in IRAS 16293-2422.
% The CS and HCN line widths range from $\sim$0.9 km s$^{-1}$ (L1551 IRS 5) to $\sim$2.6 km s$^{-1}$ (L1551 NE)
% and from $\sim$0.7 km s$^{-1}$ (B335) and $\sim$3.0 km s$^{-1}$ (L723), respectively.
The line widths of these lines are typically $\sim$1 - 2 km s$^{-1}$.
Toward L483 and L723 the HCN line widths are significantly wider than the CS line widths.
The HCN ($J$ = 4--3) line consists of six hyperfine components, two of which
($F$ = 3--3 and 4--4) can be separated from the main line ($F$ = 4--3)
by 1.977 MHz (-1.67 km s$^{-1}$) and -1.610 MHz (1.36 km s$^{-1}$), respectively
(\cite{jew97}). The intensity ratio
between the main and the other two hyperfine lines at the local thermal equilibrium
condition is 0.0217 (\cite{jew97}), and these weaker hyperfine lines are unlikely to be detectable
with the present observations. It is still possible, however, that the broader HCN
line widths than the CS line widths found in L483 and L723 are due to the presence of the hyperfine lines,
if there is any hyperfine anomaly.
% As discussed in $\S$4.2., velocity gradients in the HCN (4--3) mapping data
% are less obvious than those in the CS (7--6) mapping data, which could also be due to the presence of the hyperfine
% components.
% and the intensities of the $F$ = 3--3 and 4--4 lines
% are significantly larger than the statistical values in L483 and L723.

Toward L1551 IRS 5 ($L_{bol}$ = 22 $\LO$) and L483 ($L_{bol}$ = 13 $\LO$), the submillimeter intensities
are $\sim$twice as high as those toward the other, less bright ($L_{bol}$ $\lesssim$4.2 $\LO$) sources,
with an exception of the CS line toward B335 and the HCN line toward L43. Toward IRAS 16293-2422 ($L_{bol}$ = 21 $\LO$)
the submillimeter CS and HCN intensities are much stronger than those toward the other sources
(Fig. \ref{fig:I16spec}). These results suggest that the intensities of the submillimeter
molecular lines are related to the protostellar luminosities. We will discuss this point in $\S$4.1.

\begin{table}
\footnotesize
\begin{center}
\caption{Observed CS ($J$ = 7--6) and HCN ($J$ = 4--3) line parameters
toward the protostellar positions}\label{tab:spectra}
\begin{tabular}{lccccccccccc}
\hline\hline
 &\multicolumn{5}{c}{CS ($J$ = 7--6)} & &\multicolumn{5}{c}{HCN ($J$ = 4--3)}\\
% \colhead{} & \multicolumn{5}{c}{CS ($J$ = 7--6)} & \colhead{} & \multicolumn{5}{c}{HCN ($J$ = 4--3)} \\
% \endfirsthead
\cline{2-6} \cline{8-12}
% \hline
Source &$\int T_{MB} dv$\footnotemark[$*$] &$T_{MB}$\footnotemark[$*$] &rms &$\Delta v$\footnotemark[$*$] &$v_{\rm LSR}$\footnotemark[$*$] &
&$\int T_{MB} dv$\footnotemark[$*$] &$T_{MB}$\footnotemark[$*$] &rms &$\Delta v$\footnotemark[$*$] &$v_{\rm LSR}$\footnotemark[$*$]\\
  &(K km s$^{-1}$) &(K) &(K) &(km s$^{-1}$) &(km s$^{-1}$) & &(K km s$^{-1}$) &(K) &(K) &(km s$^{-1}$) &(km s$^{-1}$)\\
  \hline
% \colhead{Source} & \colhead{$\int T_{B} dv$\tablenotemark{a}} & \colhead{$T_{B}$\tablenotemark{a}} & \colhead{rms} & \colhead{$\Delta v$\tablenotemark{a}} & \colhead{$v_{\rm LSR}$\tablenotemark{a}} & \colhead{}
%                 & \colhead{$\int T_{B} dv$\tablenotemark{a}} & \colhead{$T_{B}$\tablenotemark{a}} & \colhead{rms} & \colhead{$\Delta v$\tablenotemark{a}} & \colhead{$v_{\rm LSR}$\tablenotemark{a}} \\
% \colhead{} & \colhead{(K km s$^{-1}$)} & \colhead{(K)} & \colhead{(K)} & \colhead{(km s$^{-1}$)} & \colhead{(km s$^{-1}$)} & \colhead{}
%           & \colhead{(K km s$^{-1}$)} & \colhead{(K)} & \colhead{(K)} & \colhead{(km s$^{-1}$)} & \colhead{(km s$^{-1}$)}}
L1551 IRS 5      &2.12\footnotemark[$\dagger$]  &1.61\footnotemark[$\dagger$] &0.10  &0.92\footnotemark[$\dagger$] &6.45\footnotemark[$\dagger$]  & &1.70\footnotemark[$\dagger$]    &1.32\footnotemark[$\dagger$] &0.12 &0.88\footnotemark[$\dagger$]   &6.29\footnotemark[$\dagger$] \\
L1551 NE        &1.26  &0.46    &0.19  &2.57 &7.18  & &- &- &0.20 &- &- \\
IRAS 16293-2422 &31.05\footnotemark[$\dagger$] &9.12\footnotemark[$\dagger$] &0.12  &2.68\footnotemark[$\dagger$] &3.56\footnotemark[$\dagger$] & &19.41\footnotemark[$\ddagger$]   &5.79\footnotemark[$\ddagger$] &0.08 &-\footnotemark[$\ddagger$]  &3.48\footnotemark[$\ddagger$] \\
L43             &0.68\footnotemark[$\S$]  &0.52\footnotemark[$\S$]  &0.18  &0.44\footnotemark[$\S$]  &0.11\footnotemark[$\S$]   & &0.85    &1.06    &0.19 &0.75    &0.41 \\
L483            &2.61  &1.80    &0.22  &1.36 &5.52  & &1.87    &0.84    &0.21 &2.09    &5.62 \\
L723            &1.21  &0.74    &0.12  &1.54 &11.24 & &1.18    &0.36    &0.09 &3.07    &11.09 \\
B335            &1.91  &1.20    &0.15  &1.49 &8.18  & &0.43    &0.58    &0.14 &0.69    &8.06\\
\hline
       \multicolumn{12}{@{}l@{}}{\hbox to 0pt{\parbox{180mm}{\footnotesize 
       \par\noindent
       \footnotemark[$*$]Derived from the one-component Gaussian fitting to
the spectra unless otherwise noted. Here, $T_{MB}$, $\Delta v$, and $v_{\rm LSR}$
are the peak main-beam brightness temperature,
FWHM width, and the central LSR velocity of the Gaussian, respectively. $\int T_{MB} dv$ 
is the integrated value of the Gaussian.
       \par\noindent
       \footnotemark[$\dagger$]Non-Gaussian spectral shapes. $T_{MB}$, $\Delta v$, $v_{\rm LSR}$, and
$\int T_{MB} dv$ are the peak main-beam brightness temperature, FWHM line width, velocity of the emission peak,
and the integrated intensity over the velocity range from $v_{\rm LSR}$ = 4.0 to 9.0 km s$^{-1}$ for L1551 IRS 5 and
-7.0 to 13.0 km s$^{-1}$ for IRAS 16293-2422, respectively.
       \par\noindent
       \footnotemark[$\ddagger$]Self-absorbed spectral shape. $T_{MB}$, $v_{\rm LSR}$, and $\int T_{MB} dv$ are
the peak main-beam brightness temperature, velocity of the emission peak, and the integrated
intensity over the velocity range from $v_{\rm LSR}$ = -7.0 to 13.0 km s$^{-1}$, respectively.
       \par\noindent
       \footnotemark[$\S$]Marginal detection and the line parameters are less reliable.
$T_{MB}$, $\Delta v$, $v_{\rm LSR}$, and $\int T_{MB} dv$ are the peak main-beam brightness temperature, FWHM line width,
velocity of the emission peak,
and the integrated intensity over the velocity range from $v_{\rm LSR}$ = -2.0 to 1.5 km s$^{-1}$, respectively.
     }\hss}}
\end{tabular}
\end{center}
\end{table}

\subsection{Distribution of the Submillimeter CS and HCN Emission}
\subsubsection{L1551 IRS 5}

L1551 IRS 5 is the brightest ($\sim$22 $\LO$) protostellar source in Taurus, and is classified as Class I \citep{eme84,fro05}.
L1551 IRS 5 was the first protostellar source where a bipolar molecular outflow was identified \citep{sne80}.
Subsequent observations of L1551 IRS 5 in CO (1--0, 2--1, 3--2) lines have revealed structures and kinematics of the
parsec-scale molecular outflow, with the blueshifted and redshifted lobes at the
south-west and north-east of the protostar, respectively \citep{uch87,sto06,mor06}.
High angular-resolution ($\sim$0$\farcs$04) 7-mm continuum observations of L1551 IRS 5 with VLA
have found that L1551 IRS 5 is a triple protostellar system with a projected separation of $\sim$47 AU and $\sim$13 AU \citep{lim06}.
SMA observations of L1551 IRS 5 in the CO (2--1) line within the $\sim$4000 AU region
have found multiple molecular outflows,
one of which shows precessing motion presumably due to the tidal interaction of the circumstellar disks
around the multiple protostars \citep{wu09}.
Interferometric observations of L1551 IRS 5 in millimeter molecular lines of CS (2--1), H$^{13}$CO$^{+}$ (1--0),
and C$^{18}$O (1--0) have revealed a $\sim$2000 - 5000 AU-scale molecular envelope surrounding L1551 IRS 5, which is
elongated perpendicularly to the outflow axis \citep{oh96b,sai96,mom98}.
Along the outflow direction the south-western and north-eastern parts of the envelope are blueshifted and redshifted,
whereas across the outflow direction the north-western and south-eastern parts are redshifted and blueshifted, respectively.
The velocity gradients seen in the envelope along and across the outflow direction have been interpreted as
an infalling and rotating gas motion in the envelope, respectively \citep{oh96b,sai96,mom98}.
On the other hand, SMA observations of L1551 IRS 5
in the submillimeter CS (7--6) line have found a compact ($\sim$500 AU) molecular gas with mainly rotating motion
around the protostar \citep{tak04}.
With the present ASTE observations of L1551 IRS 5 we can investigate the extent and kinematics of the warm gas traced
in the submillimeter molecular lines and the relation with the infalling envelope traced in the millimeter lines.

\begin{figure}
\begin{center}
\FigureFile(160mm,160mm){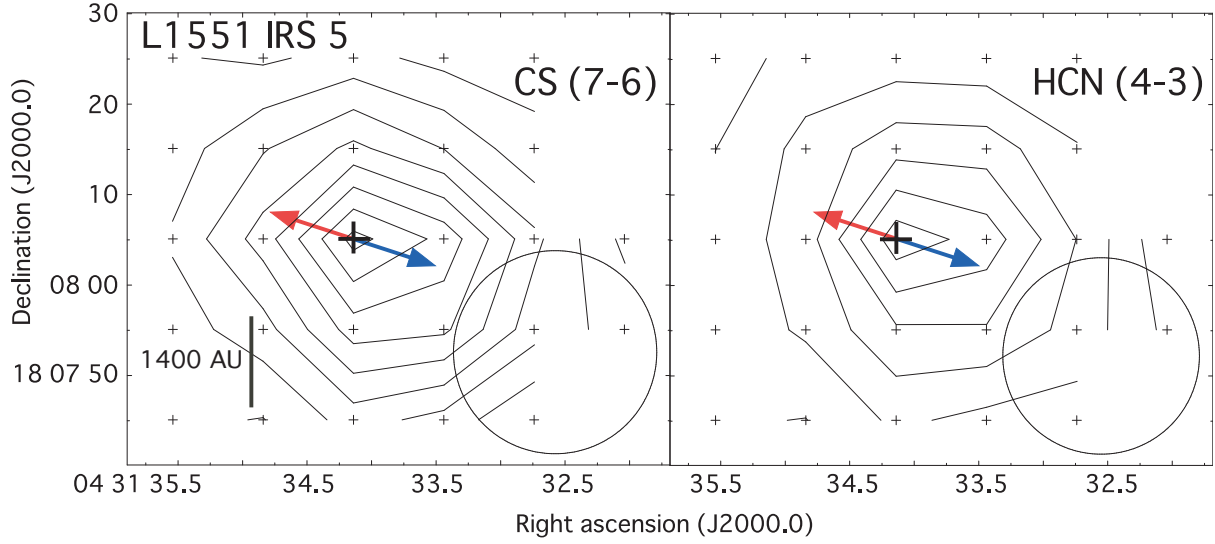}
\end{center}
\caption{Total integrated intensity maps of the CS (7--6) ($left$) and HCN (4--3) ($right$)
emission in L1551 IRS 5 over the velocity range of 5.3 - 8.0 km s$^{-1}$ observed with ASTE.
Contour levels are from 2 $\sigma$ in steps of 4 $\sigma$ (1 $\sigma$ = 0.0605 K km s$^{-1}$).
The highest contours in the CS and HCN maps are 30 $\sigma$ and 22 $\sigma$, respectively.
Crosses indicate observed positions.
Red and blue arrows show the direction of the redshifted and blueshifted
molecular outflow, respectively, and the roots of the arrows with large crosses indicate the
protostellar position. An open circle at the bottom-right corner in each panel shows the ASTE beam.
}\label{fig:irs5map}
\end{figure}

\begin{figure}
\begin{center}
\FigureFile(140mm,140mm){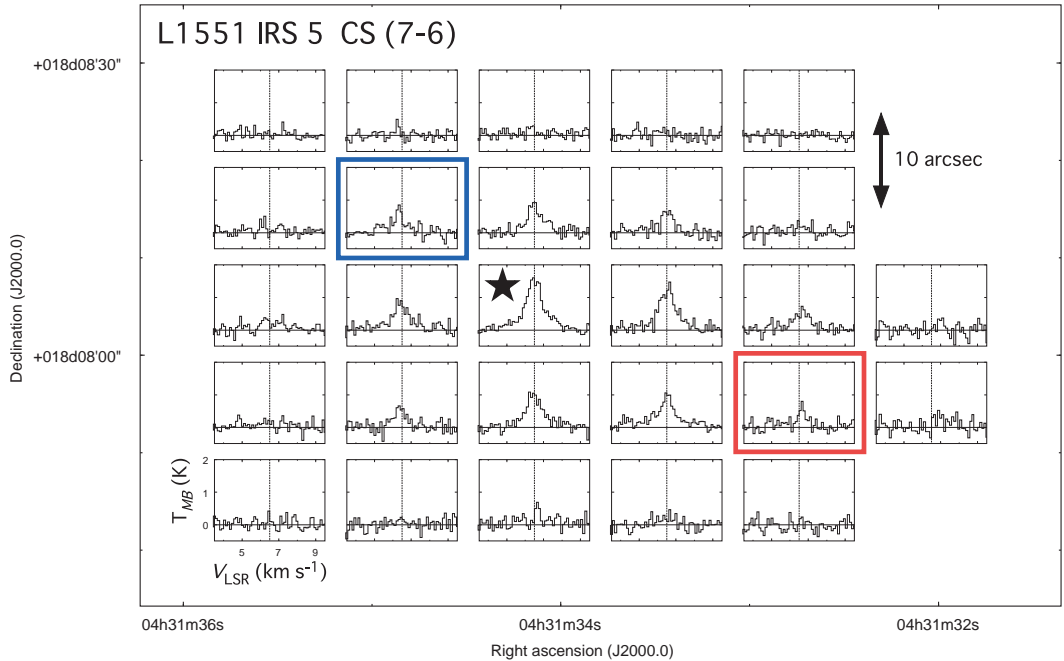}
\end{center}
\caption{Stamp map of the CS (7--6) line profile in L1551 IRS 5 observed with ASTE.
The line profile with a star mark is the profile toward the central stellar position.
Solid horizontal lines show zero levels, and dashed vertical lines systemic velocity of
6.5 km s$^{-1}$.}\label{fig:irs5cs}
\end{figure}

\begin{figure}
\begin{center}
\FigureFile(140mm,140mm){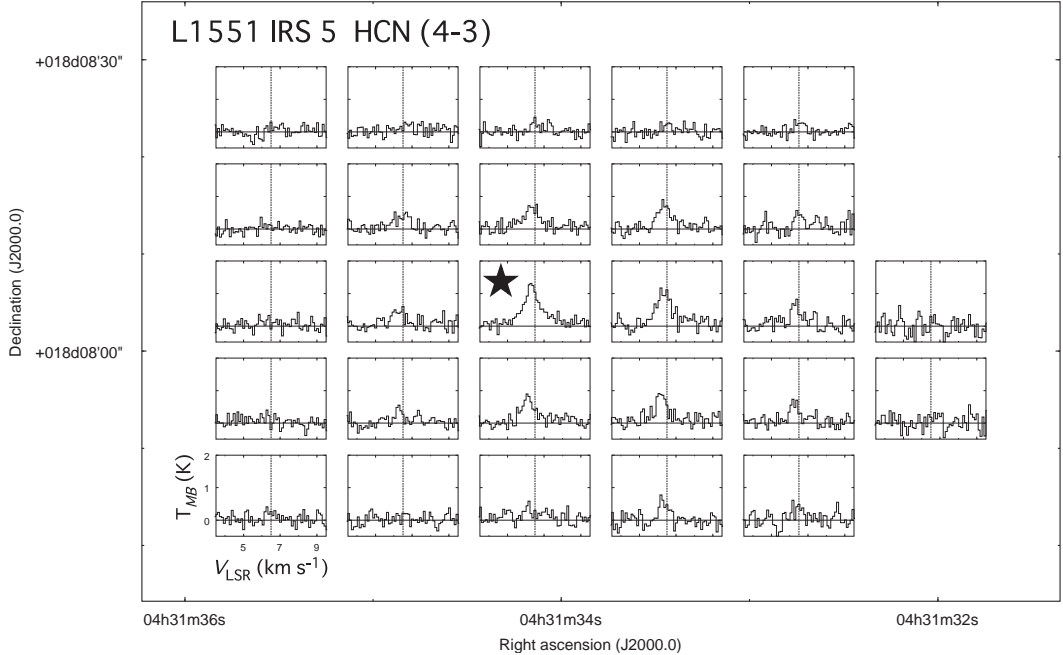}
\end{center}
\caption{Same as Figure \ref{fig:irs5cs} but for the HCN (4--3) line profile.}\label{fig:irs5hcn}
\end{figure}

\begin{figure}
\begin{center}
\FigureFile(160mm,160mm){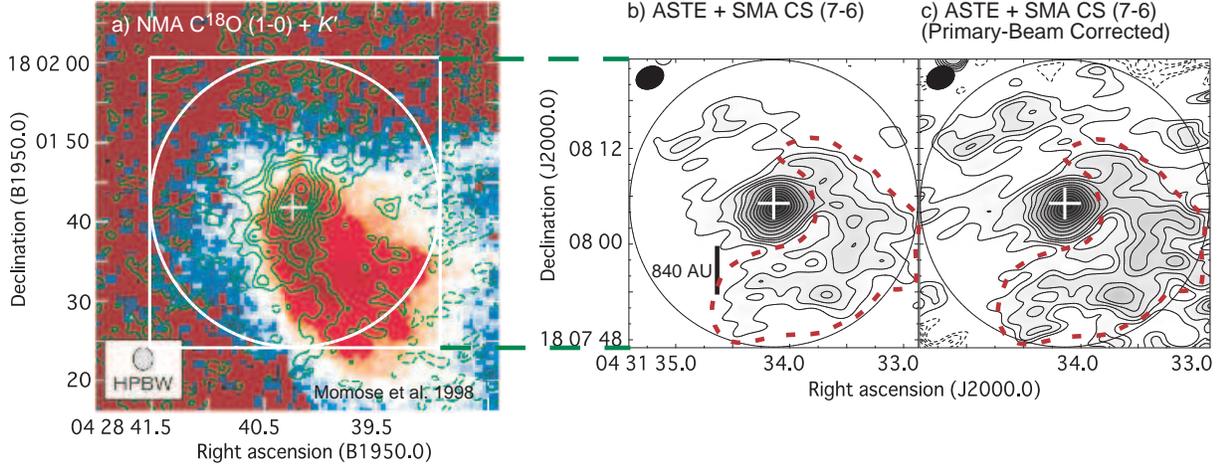}
\end{center}
\caption{a) Total integrated intensity map
of the C$^{18}$O (1--0) emission observed with NMA (green contours) superposed on
the $K^{\prime}$-band image of the associated reflection nebula in L1551 IRS 5,
taken from Fig. 2 by Momose et al. (1998). Contour levels start from 1.5 $\sigma$
in steps of 1.5 $\sigma$ (1 $\sigma$ = 0.679 K).
A white cross indicates the protostellar position, and a filled ellipse at the bottom-left corner
shows the NMA beam (2\farcs8 $\times$ 2\farcs5, P.A. = 0$^{\circ}$). An open circle and square show the SMA Field of View and the
imaging region of the SMA images shown in middle and right panels.
b) Combined ASTE + SMA image of the CS (7--6) emission
in L1551 IRS 5. Contour levels are from 2 $\sigma$ in steps of 1 $\sigma$ (1 $\sigma$ = 0.904 K km s$^{-1}$).
The integrated velocity range is from 5.4 km s$^{-1}$ to 8.4 km s$^{-1}$.
A cross and an open circle indicate the protostellar position and the SMA Field of View, respectively.
A filled ellipse at the top-left corner shows the synthesized beam (3\farcs6 $\times$ 2\farcs7, P.A. = -65$^{\circ}$).
A brown dashed contour delineates the extended, halo-like CS component (see text).
c) Combined ASTE + SMA image of the CS (7--6) emission in L1551 IRS 5 after the primary beam correction.
Contour levels and symbols are the same as those in the middle panel.
}
\label{fig:l15comb}
\end{figure}

Figure \ref{fig:irs5map} shows total integrated intensity maps of the CS ($J$ = 7--6) ($left$ $panel$) and HCN ($J$ = 4--3) ($right$)
lines in L1551 IRS 5. Figures \ref{fig:irs5cs} and \ref{fig:irs5hcn} show relevant line profile maps of the CS (7--6) and HCN (4--3) lines,
respectively. There are intense blobs centered on the protostellar position both in the CS and HCN emissions.
Both the CS and HCN emission distributions show a slight elongation toward the south-west of the protostar.
The shapes of the emission distributions do not resemble the shape of the ASTE beam, suggesting that
the submillimeter emissions are slightly resolved. Two-dimensional Gaussian fittings to the submillimeter
emission distributions show that the deconvolved size of the HCN emission is $\sim$18$\arcsec$ $\times$ 15$\arcsec$
(= 2600 AU $\times$ 2100 AU; PA = 65$^{\circ}$ $\pm$ 12$^{\circ}$), and that in the CS emission only the major
axis (PA = 70$^{\circ}$ $\pm$ 1$^{\circ}$) is resolved ($\sim$18$\arcsec$).
The deconvolved sizes are smaller than the
ASTE beam size and hence the estimated emission extents are not well-determined.
In order to unambiguously measure the extent of the submillimeter molecular emission, we combined the present ASTE CS (7--6) data
with our published SMA CS (7--6) data \citep{tak04}. The $middle$ and $right$ panels in
Figure \ref{fig:l15comb} show the
combined ASTE + SMA total integrated intensity maps of the CS (7--6) emission in L1551 IRS 5 before and after the primary beam correction,
respectively.
The submillimeter CS emission in L1551 IRS 5 consists of two components;
one is a compact ($\lesssim$500 AU) component centered on the protostellar positions, and the other
a halo-like, extended ($\sim$2000 AU) component, delineated by brown dashed curves in Figure \ref{fig:l15comb}.
The central compact CS (7--6) component is also seen in the SMA-only image, which shows rotating gas motion
inside the infalling envelope seen in the C$^{18}$O (1--0) line \citep{tak04}. The combined image
shows another submillimeter CS component, which appears to extend toward the
associated reflection nebula and hence the direction of the blueshifted molecular outflow, and
the extension of this submillimeter emission component appears to be perpendicular to that of
the C$^{18}$O (1--0) envelope (see Figure \ref{fig:l15comb} $left$).
The ``asymmetric'', elongated distribution of the submillimeter molecular emissions
toward the direction of the blueshifted molecular outflow is also seen in L483 (Paper I).

The CS (7--6) line profile map shows that the emission peaks at the southwest of the protostar tend to be redshifted
and at the northeast blueshifted (Figure \ref{fig:irs5cs}, see spectra with blue and red rectangles),
although in the HCN (4--3) line profile map such a velocity structure is not clear (Figure \ref{fig:irs5hcn}).
This velocity feature in the submillimeter CS line appears to be
opposite to that of the infalling envelope seen in the millimeter C$^{18}$O and H$^{13}$CO$^{+}$ (1--0) lines \citep{sai96,mom98},
and that of the CO outflows \citep{sto06,mor06,wu09}.
Opposite velocity gradients of the submillimeter molecular lines from those of the millimeter lines
and the associated outflows are also seen in L483 and B335 (Paper I).
In $\S$4.2., we will present quantitative analyses of the submillimeter velocity gradients in these protostellar sources
and discuss their origin.

% \begin{figure}
% \begin{center}
% \FigureFile(120mm,120mm){L1551PV.eps}
% \end{center}
% \caption{Position-Velocity (P-V) diagrams of the CS (7--6) (left) and HCN (4--3) lines (right)
% across (upper) and along (lower) the axis of the associated molecular outflow (P.A. = 72$^{\circ}$)
% passing through the central stellar position in L1551 IRS 5. Contour levels are from
% 2 $\sigma$ in steps of 2 $\sigma$ (1 $\sigma$ = 0.11 K). Dashed lines delineate
% detected velocity gradients.}\label{fig:irs5pv}
% \end{figure}

\subsubsection{L1551 NE}

L1551 NE is the second brightest ($\sim$4.2 $\LO$) protostellar source in Taurus,
located $\sim$2$\farcm$5 north-east of L1551 IRS 5 \citep{eme84}.
L1551 NE is classified as a Class I protostar based on its spectral energy distribution \citep{fro05}.
Radio continuum observations by \citet{rod95} and \citet{rei02} have revealed
two compact sources with a separation of $\sim$70 AU and a position angle of $\sim$300$^{\circ}$,
suggesting that L1551 NE is a protostellar binary.
Optical and NIR observations of L1551 NE have found Herbig-Haro objects
HH 28, HH 29, HH 454, and a collimated [Fe~II] jet,
and a bright reflection nebula \citep{dra85,hod95,dev99,rei00,hay09}.
With JCMT, \citet{mo95a} detected blueshifted ($\sim$3 - 5 km s$^{-1}$ from the systemic velocity) emission
in the wing of the CO (3--2) line toward L1551 NE that coincides spatially with its
optical/infrared reflection nebula, suggesting that the blueshifted emission traces the molecular outflow
driven by L1551 NE.
Millimeter interferometric observations of L1551 NE in the CS (2--1, 3--2) lines revealed
a compact ($\sim$700 AU) dense-gas component associated with the central protostar
plus a NW-SE elongated ($\sim$6000 AU) component located toward $\sim$1000 AU
west from the protostar \citep{yok03}. On the other hand, interferometric observations of the H$^{13}$CO$^{+}$ (1--0) line show a
NW-SE elongated ($\sim$8000 AU) component located toward $\sim$700 AU
east of the protostar \citep{sai01}.
The prominent redshifted molecular outflow from L1551 IRS5 passes in projection
through the location of L1551~NE, and likely impacts the molecular envelope around L1551 NE.
The anti-correlation between the CS and H$^{13}$CO$^{+}$ emissions may be due to
the shock chemistry \citep{pla95,yok03}.

% 1sigma = 0.0477 K km s-1
% CS  
% 0.40078, 0.63436, 0.77248, 0.59385 K km s-1
% 8.4 sigma, 13.3 sigma, 16.2 sigma, 12.4 sigma
%
% HCN
% 0.14013, 0.22195, 0.44310, 0.30976 K km s-1
% 2.9 sigma, 4.7 sigma, 9.3 sigma, 6.5 sigma

\begin{figure}
\begin{center}
\FigureFile(140mm,140mm){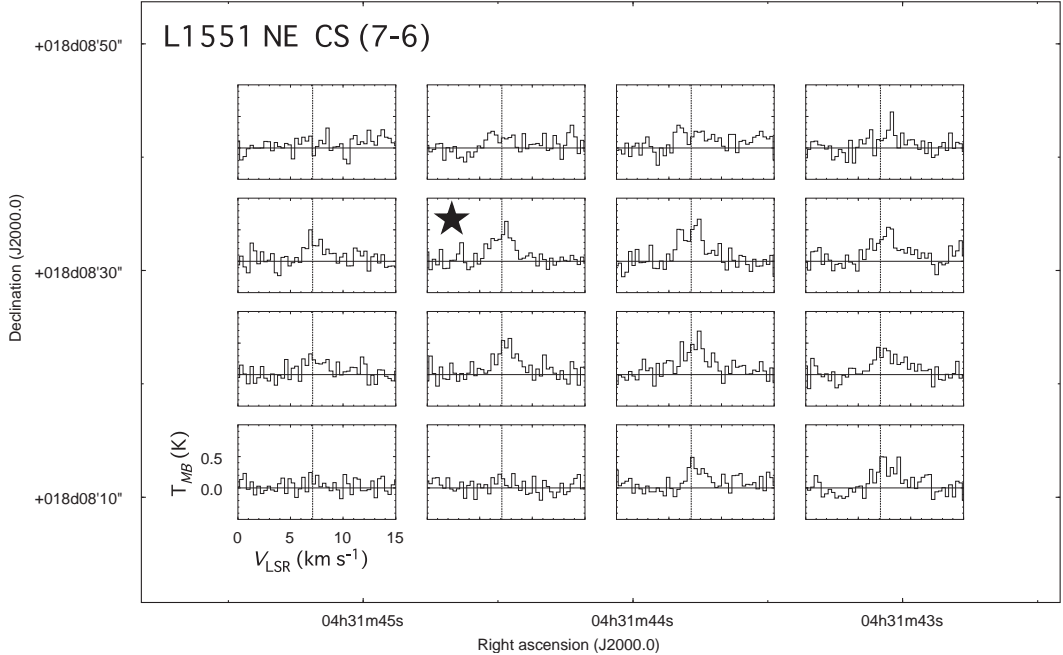}
\end{center}
\caption{Stamp map of the CS (7--6) line profile in L1551 NE observed with ASTE.
The line profile with a star mark is the profile toward the central stellar position.
Solid horizontal lines show zero levels, and dashed vertical lines systemic velocity of
7.1 km s$^{-1}$. Three-channel bindings were performed to increase the signal-to-noise
ratio of the spectra.}\label{fig:necs}
\end{figure}

\begin{figure}
\begin{center}
\FigureFile(140mm,140mm){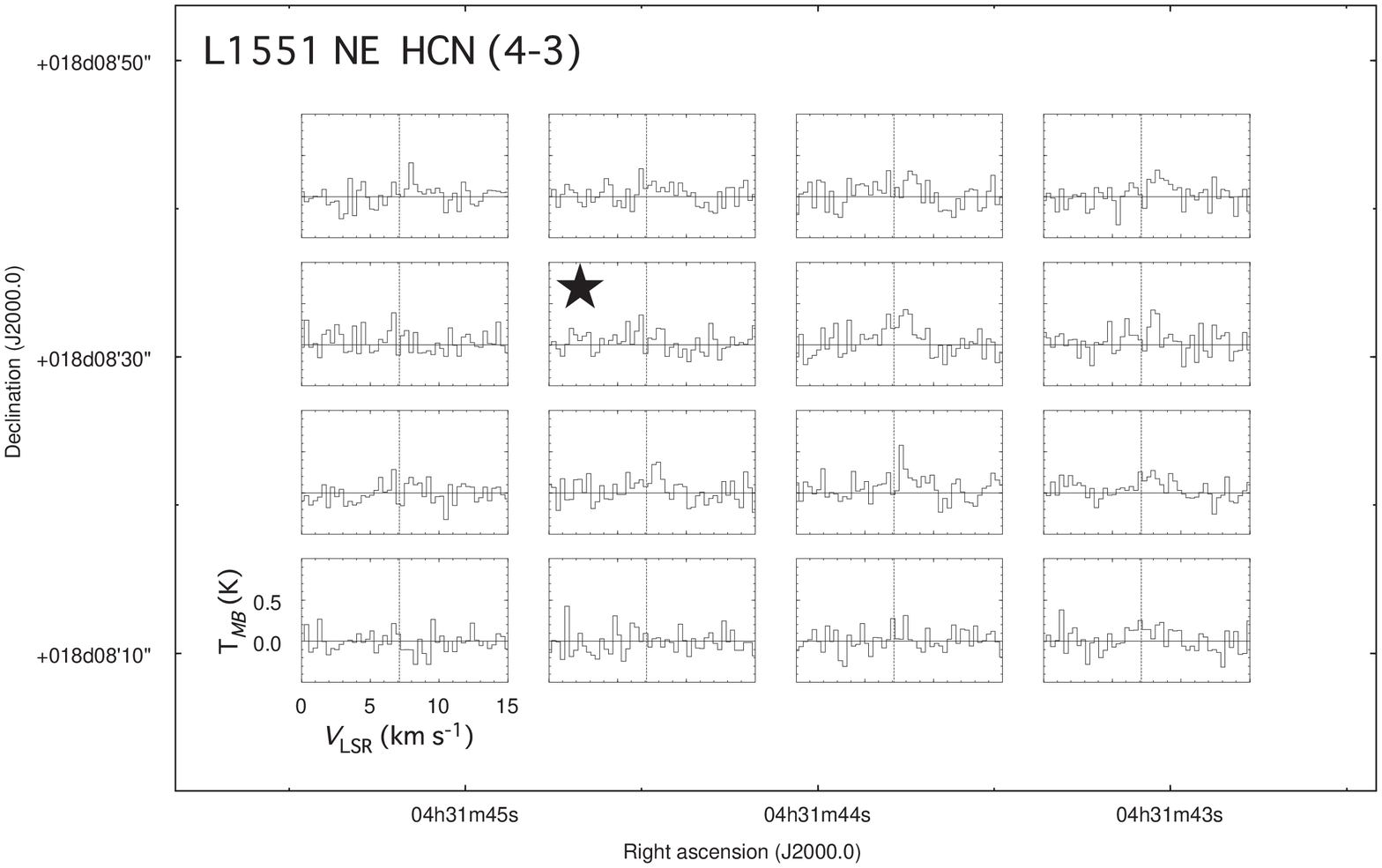}
\end{center}
\caption{Same as Figure  \ref{fig:necs} but for the HCN (4--3) emission.}\label{fig:nehcn}
\end{figure}

\begin{figure}
\begin{center}
\FigureFile(140mm,140mm){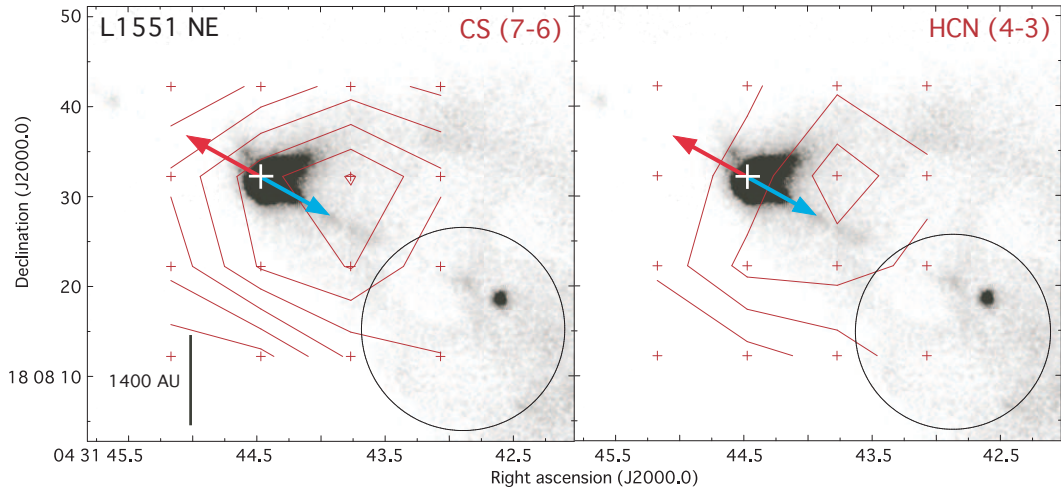}
\end{center}
\caption{Total integrated intensity maps of the CS (7--6) ($left$) and HCN (4--3) ($right$) emission
in L1551 NE over the velocity range of 5.6 - 8.7 km s$^{-1}$
observed with ASTE, superposed on the [Fe II] + 1.64-$\micron$
continuum image taken with SUBARU in grey scale (Hayashi \& Pyo 2009;
The large black blob associated with the protostar is the reflection nebula,
and the small blob at the south-west of the protostar is HH 454A.).
Contour levels are from 2 $\sigma$ in steps of 2 $\sigma$ (1 $\sigma$ = 0.080 K km s$^{-1}$).
The highest contours in the CS and HCN maps are 16 $\sigma$ and 8 $\sigma$, respectively.
Crosses indicate observed positions.
Red and blue arrows show the direction of the redshifted and blueshifted
molecular outflow, respectively, and the roots of the arrows with white crosses indicate the
protostellar position.
An open circle at the bottom-right corner in each panel shows the ASTE beam.
}\label{fig:nemap}
\end{figure}

Figures \ref{fig:necs} and \ref{fig:nehcn} show line profile maps of the CS (7--6) and HCN (4--3) emissions
in L1551 NE, respectively. Figure \ref{fig:nemap} shows the relevant total integrated intensity maps of the
CS (7--6) ($left$ $panel$) and HCN (4--3) ($right$) lines in L1551 NE (red contours), superposed on the
[Fe II] + 1.64-$\micron$ continuum image taken with SUBARU \citep{hay09}.
Although the submillimeter molecular emissions are not spatially resolved with the ASTE
22$\arcsec$ beam, their peak positions appear to be $\sim$10$\arcsec$ west from the protostellar position.
Along the east-west direction passing through the central protostellar position,
the CS integrated intensity from -20$\arcsec$ to +10$\arcsec$ of the protostar
exceeds above the 8$\sigma$ level, and the HCN integrated intensity from -10$\arcsec$ to +10$\arcsec$
above 4$\sigma$ level.
Taking into account the detection levels, the ASTE pointing error ($\sim$2$\arcsec$) and the beam size,
we consider that these $\sim$10$\arcsec$ offsets of the
submillimeter emission peaks from the protostellar position are real.
% Along the east-west direction passing through the central protostellar position, the CS total integrated intensities
% toward +10$\arcsec$, 0$\arcsec$, -10$\arcsec$, and -20$\arcsec$ of the protostellar position
% are 8.4 $\sigma$, 13.3 $\sigma$, 16.2 $\sigma$, and 12.4 $\sigma$, respectively.
% Likewise, the HCN intensities toward +10$\arcsec$, 0$\arcsec$, -10$\arcsec$, and -20$\arcsec$
% of the protostellar position are 2.9 $\sigma$, 4.7 $\sigma$, 9.3 $\sigma$, and 6.5 $\sigma$, respectively.
% Taking into account these emission distributions with respect to the noise level, the ASTE pointing error
% ($\sim$2$\arcsec$) and the beam size, we consider that these $\sim$10$\arcsec$ offsets of the
% submillimeter emission peaks from the protostellar position are real.
These offsets of the emission distributions
resemble those of the millimeter CS (2--1, 3--2) emission distributions \citep{yok03}.
We also note that the offsets of the submillimeter molecular emissions are toward the direction
of the associated reflection nebula seen in the SUBARU image, which is similar to the case
of L1551 IRS 5 (Fig. \ref{fig:l15comb}).

\subsubsection{L723 and L43}

L723 is another Class 0 candidate with a luminosity of $\sim$3.3 $\LO$ at a distance of 300 pc \citep{dav87}.
In L723 there are at least four centimeter sources; VLA 1, and VLA 2A, 2B, and 2C, and one 7-mm
source; VLA 2D \citep{ang91,car08}.
VLA 1 is located at $\sim$4500 AU south-west from the VLA 2 complex, and only the VLA2 complex
is associated with dense molecular gas seen in the millimeter CS (1--0, 2--1, 3--2) lines
as well as the $X$-shaped CO outflow \citep{hay91,hir98}. The peaks of the CS (2--1, 3--2) emission
distributions are offset by $\sim$10$\arcsec$ east from the VLA 2 complex \citep{hir98}.
% VLA 2A and 2B comprise
% a closely-separated ($\sim$ 90 AU) protostellar binary, while VLA 2C is located
% $\sim$200 AU northeast of VLA 2A \citep{car08}. There is another 7-mm source, VLA 2D, located
% $\sim$1000 AU southeast of VLA 2A \citep{car08}.
SMA observations of L723 have found two compact ($\sim$600 AU) dusty cores at a
projected separation of $\sim$880 AU (SMA 1 and SMA 2), where SMA 1 is associated with VLA 2D
and SMA 2 VLA 2A, 2B, and 2C \citep{gir09}. Both SMA 1 and 2 show marginal evidence of infalling and rotating
gas motion in the H$_{2}$CO (3$_{0,3}$--2$_{0,2}$) line, whereas only SMA 2 appears to be associated with
HH objects and multiple (at least three) molecular outflows driven by the three different protostellar sources
\citep{car08,gir09}.
% which comprise the large-scale $X$-shaped CO outflow

% L723 total integrated
% 1 sigma = 0.0433 K km s-1
%            +20 +10, 0 -10, VLA 1
% CS
% 0.19065, 0.60902,                    0.58955, 0.26863, 0.092116 K km s-1
%  4.40 sigma, 14.065 sigma  13.615 sigma, 6.204 sigma, 2.13 sigma
% HCN
% 0.26765, 0.40992, 0.58540,                     0.25662, 0.11078 K km s-1
%  6.18 sigma, 9.47 sigma, 13.52 sigma,  5.93 sigma, 2.56 sigma

\begin{figure}
\begin{center}
\FigureFile(140mm,140mm){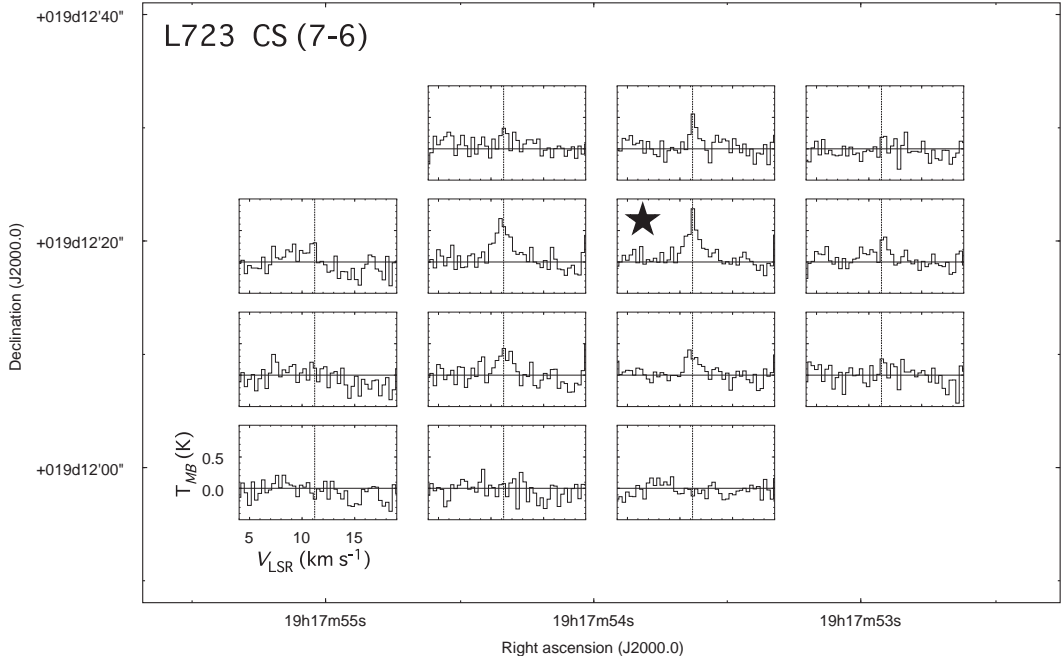}
\end{center}
\caption{Stamp map of the CS (7--6) line profile in L723 observed with ASTE.
The line profile with a star mark is the profile toward the central stellar position.
Solid horizontal lines show zero levels, and dashed vertical lines systemic velocity of
11.2 km s$^{-1}$. Three-channel bindings were performed to increase the signal-to-noise
ratio of the spectra.}\label{fig:l723cs}
\end{figure}

\begin{figure}
\begin{center}
\FigureFile(140mm,140mm){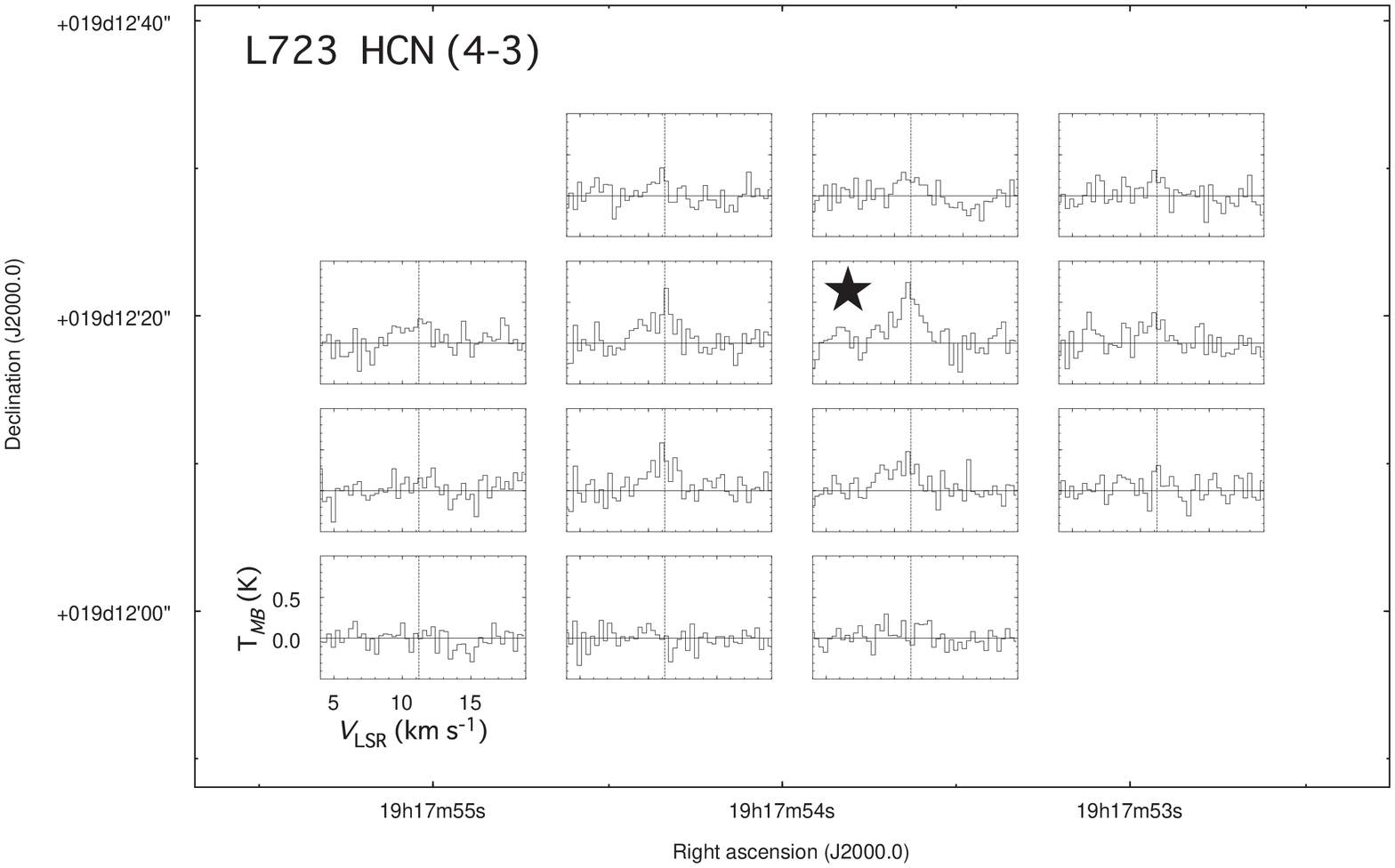}
\end{center}
\caption{Same as Figure \ref{fig:l723cs} but for the HCN (4--3) emission.}\label{fig:l723hcn}
\end{figure}

\begin{figure}
\begin{center}
\FigureFile(120mm,120mm){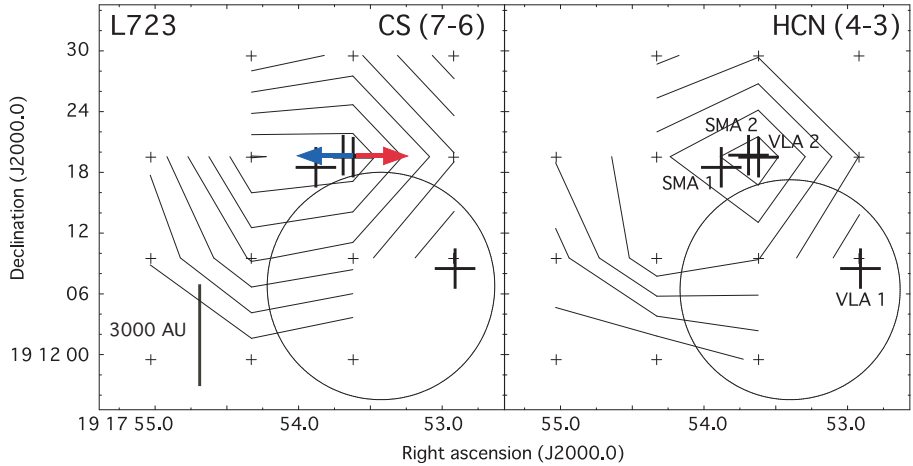}
\end{center}
\caption{Total integrated intensity maps of the CS (7--6) ($left$) and HCN (4--3) ($right$)
emission in L723 over the velocity range of 9.9 - 12.1  km s$^{-1}$
observed with ASTE.
Contour levels are from 2 $\sigma$ in steps of 2 $\sigma$ (1 $\sigma$ = 0.072 K km s$^{-1}$).
The highest contours in the CS and HCN maps are 14 $\sigma$ and 12 $\sigma$, respectively.
Small crosses indicate observed positions, and large crosses positions
of VLA 1, VLA 2 (A, B, C), SMA 1, and SMA 2, as labeled in the figure.
Red and blue arrows show the direction of the redshifted and blueshifted
molecular outflow driven from VLA 2, respectively.
An open circle at the bottom-right corner in each panel shows the ASTE beam.
}\label{fig:l723map}
\end{figure}

Figures \ref{fig:l723cs}, \ref{fig:l723hcn}, and \ref{fig:l723map} show line profile maps of the
CS (7--6) and HCN (4--3) emission, and the corresponding total integrated intensity maps in L723, respectively.
The CS and HCN total integrated intensities toward VLA 2 exceed $>$ 13$\sigma$,
whereas those toward VLA 1 are below 3$\sigma$.
These results suggest that only VLA 2 is associated with the dense and/or warm molecular gas traced by the
submillimeter molecular lines, which is consistent with the previous millimeter observations \citep{hir98,gir09}.
The intensity of the CS emission at 10$\arcsec$ east of the protostar
% (14.1 $\sigma$)
is comparable to that at the protostellar position ($\sim$14 $\sigma$),
and is much higher than that at 10$\arcsec$ west ($\sim$6 $\sigma$).
The HCN intensity at 10$\arcsec$ east ($\sim$10 $\sigma$) is also higher than that at 10$\arcsec$ west ($\sim$6 $\sigma$).
Although the ASTE observations do not resolve the submillimeter emissions in L723,
the differences of the submillimeter intensities between 10$\arcsec$ east and west of the protostar are probably real. Hence, the peaks of the
submillimeter emission distributions in L723 are likely offset from the protostellar position, as in the case of L1551 NE.
The millimeter CS (2--1, 3--2) emission distributions also show similar peak offsets by $\sim$10$\arcsec$ east
from the protostar \citep{hir98}. The direction of these offsets is toward the associated blueshifted outflow.

% \begin{figure}
% \begin{center}
% \FigureFile(120mm,120mm){L483HCNTafalla.eps}
% \end{center}
% \caption{Total integrated intensity map of the HCN (4--3) emission in L483
% shown in Paper I (black contours), superposed on the contour map of the high-velocity
% $^{12}$CO (2--1) emission (white contours) and K$^{\prime}$ image taken from Figure 2 of
% Tafalla et al. (2000).}\label{fig:l483map}
% \end{figure}

\begin{figure}
\begin{center}
\FigureFile(140mm,140mm){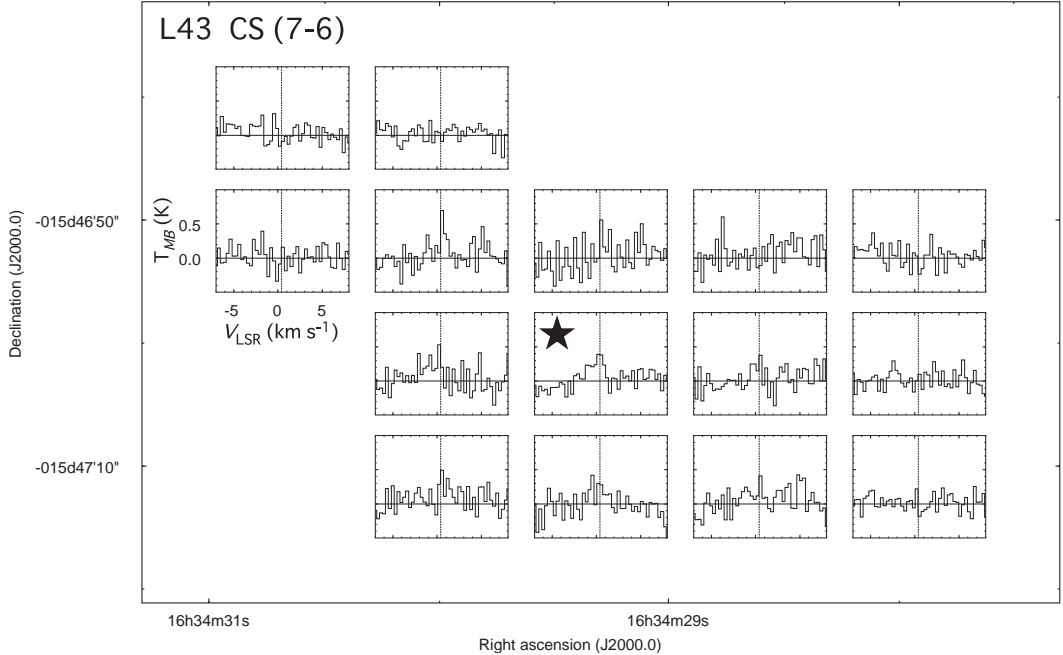}
\end{center}
\caption{Stamp map of the CS (7--6) line profile in L43 observed with ASTE.
The line profile with a star mark is the profile toward the central stellar position.
Solid horizontal lines show zero levels, and dashed vertical lines systemic velocity of
0.4 km s$^{-1}$. Three-channel bindings were performed to increase the signal-to-noise
ratio of the spectra.}\label{fig:l43cs}
\end{figure}

\begin{figure}
\begin{center}
\FigureFile(140mm,140mm){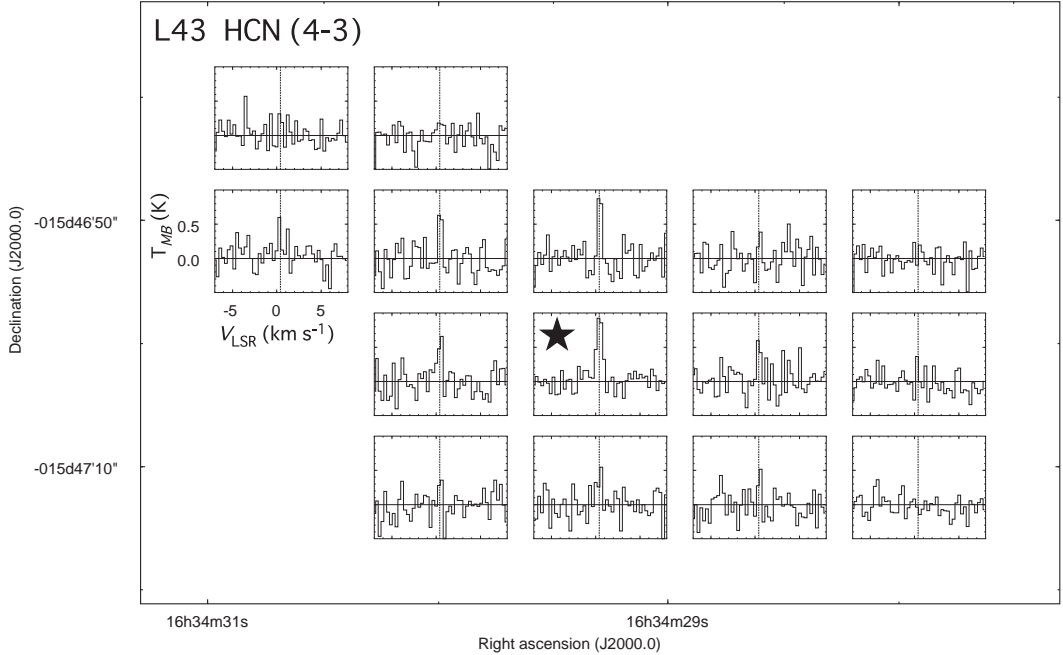}
\end{center}
\caption{Same as Figure \ref{fig:l43cs} but for the HCN (4--3) emission.}\label{fig:l43hcn}
\end{figure}

Figures \ref{fig:l43cs} and \ref{fig:l43hcn} show line profile maps of the CS (7--6) and HCN (4--3) emission
in L43, respectively.
L43 ($or$ RNO 91, IRAS 16316-1540) is a Class I - II protostar with $L_{bol}$ $\sim$2.7 $\LO$
and $T_{bol}$ $\sim$370 K \citep{shi00,ang02,che09}, and hence somewhat a more evolved source
than the other sources of our sample.
The protostar is associated with an $\sim$4000-AU scale dusty
envelope \citep{shi00,che09}, and a bipolar molecular outflow along the SE - NW direction \citep{ben98,lee00,lee02}.
% A highly-asymmetric bipolar molecular outflow has been
% found in L43, with the blueshifted lobe at the southeast much stronger than the redshifted lobe
% at the northwest of the protostar \citep{ben98,lee00}. The blueshifted
% outflow lobe shows a peculiar $U$-shaped morphology \citep{ben98,lee00,lee02}.
Figures \ref{fig:l43cs} and \ref{fig:l43hcn} show that
the signal-to-noise ratio of the submillimeter molecular lines is not high enough to discuss their spatial distributions.
One interesting point is that L43 is the only source in our ASTE observations with the HCN (4--3) emission
more intense than the CS (7--6) emission. This could be related to the later protostellar evolutionary stage of L43 than that of
the other sources.

\section{Discussion}
\subsection{``Skewed'' Distributions of the Submillimeter Molecular Emissions}

As described in the previous section, the submillimeter CS (7--6) and HCN (4--3) emissions
toward low-mass protostars often show ``skewed'' distributions with respect to the protostellar positions.
The combined ASTE + SMA image of the CS emission in L1551 IRS 5
illustrates that there is a halo-like component elongated ($\sim$2000 AU) toward the associated
reflection nebula and the blueshifted outflow, as well as a central compact ($\lesssim$500 AU)
component centered on the protostellar position.
In L1551 NE, the emission peaks of the CS and HCN lines are likely offset from the protostellar position,
and are shifted toward the associated reflection nebula.
Similar peak offsets of the submillimeter CS and HCN emissions are also seen in L723.
In L483, we found possible extensions of the CS ($\sim$2300 AU) and HCN ($\sim$5500 AU) emissions
toward the direction of the associated reflection nebula and the blueshifted outflow (Paper I).
The combined ASTE + SMA CS (7--6) image in B335, another Class 0 protostar, also shows
an elongated ($\sim$2000 AU) component tracing the reflection nebula as well as a
compact ($\lesssim$500 AU) component centered on the protostellar position \citep{yen11}.
These results indicate that the asymmetric, or ``skewed'' distributions of the submillimeter molecular
emissions toward the associated reflection nebulae are relatively frequent.

\begin{figure}
\begin{center}
\FigureFile(100mm,100mm){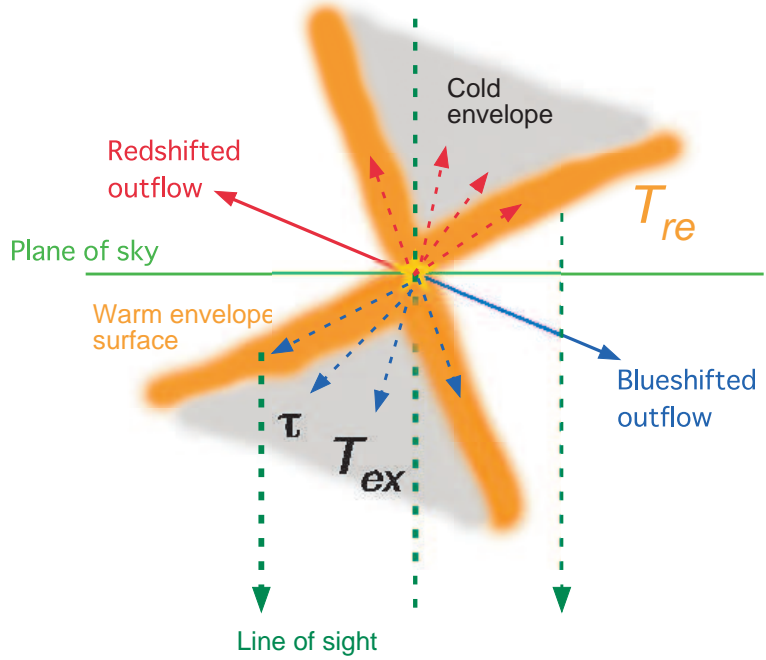}
\end{center}
\caption{Schematic diagram of the envelope configuration to explain the observed skewed distribution and the velocity
gradient of the submillimeter molecular emissions.
The orange area denotes the region of the warm surface of the cavity wall in the envelope, where the submillimeter molecular emissions arise.
The gray area indicates the cold inner envelope.
The star mark shows the central protostar, and the blue and red arrows the associated blueshifted and redshifted outflows,
respectively. The horizontal green line shows the plane of the sky, whereas the vertical dashed arrows the line of sight.
$T_{re}$, $T_{ex}$, and $\tau$ denote the brightness temperature of the submillimeter molecular emission at the cavity wall,
the excitation temperature and the optical depth of the submillimeter emission in the cold inner envelope,
respectively. In reality, the cavity wall is also present at the foreground and the background of the paper,
and the dashed arrows denote the dispersing gas motion at the cavity wall.
This figure illustrates that at the side of the blueshifted outflow the submillimeter molecular emission from
the warm cavity surface is seen directly whereas at the side of the redshifted outflow the emission is absorbed by
the cold foreground envelope component. Furthermore,
at the side of the blueshifted outflow the envelope dispersing motion
is mostly observed as a redshifted submillimeter emission whereas at the side of the redshifted outflow as a blueshifted
submillimeter emission.
}\label{fig:envconf}
\end{figure}

One possible interpretation of these results is that the extended submillimeter molecular emissions
at a few thousands AU scale trace ``walls'' of the cavity of the envelope evacuated by the associated outflows.
In Figure \ref{fig:envconf} we show a schematic diagram to explain the skewed distribution of
the submillimeter molecular emissions.
At the surface of the cavity wall molecular gas is irradiated by the stellar photons and hence the gas
temperature is likely to be higher than that at the midplane \citep{nak03,whi03}.
As discussed in Paper I, the submillimeter CS and HCN emissions likely trace warmer ($\gtrsim$40 K) molecular gas
than that traced by millimeter molecular emissions, and hence the warmer cavity walls are selectively traced 
by the submillimeter molecular emissions whereas the colder, midplane regions are traced 
by the millimeter emissions.
Here, at the side of the associated reflection nebula the wall is directly seen along the line of sight,
while at the other side the wall is likely obscured by the cold foreground envelope gas.
The skewness of the submillimeter molecular emissions, where the submillimeter emissions are
bright only at the side of the associated reflection nebula, may be explained by the foreground absorption
along the line of sight.
% as in the case of the one-side morphology of the reflection nebula.

\begin{figure}
\begin{center}
\FigureFile(70mm,70mm){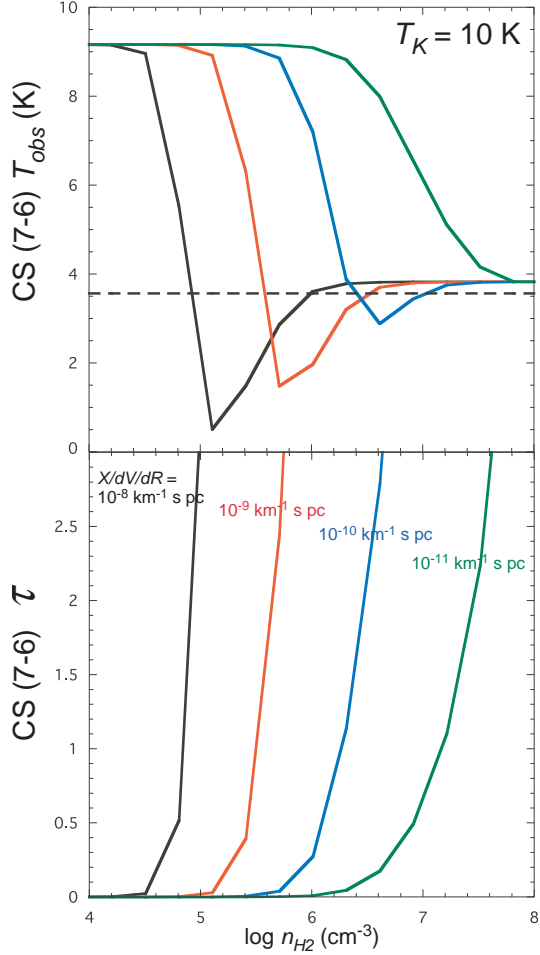}
\end{center}
\caption{Result of our statistical equilibrium calculations of the CS (7--6) emission. The background CS (7--6) intensity
is fixed to be 9.2 K, the observed peak intensity at the south-western extended CS (7--6) emission component in L1551 IRS 5.
We calculated the excitation temperature and the optical depth of the CS (7--6) emission in the foreground cloud component
with the LVG model, and then calculated the CS intensity along the line of sight. Upper and lower panels show the calculated
CS (7--6) brightness temperature and the optical depth, respectively,
as a function of the molecular-gas density in the foreground cloud with a kinetic temperature
of 10 K, and $X/dV/dR$ values of 10$^{-8}$ km$^{-1}$ s pc (black curve), 10$^{-9}$ km$^{-1}$ s pc (red),
10$^{-10}$ km$^{-1}$ s pc (blue), and 10$^{-11}$ km$^{-1}$ s pc (green). A horizontal dashed line in the upper panel
shows the 3 $\sigma$ upper limit of the combined ASTE + SMA CS (7--6) data. This figure illustrates that under the presence of
cold dense foreground molecular gas the CS (7--6) emission can be significantly absorbed.}\label{fig:lvg}
\end{figure}

In order to illustrate this possibility, we performed simple calculations based on the
Large Velocity Gradient (LVG) model \citep{gol74,sco74}. The brightness temperature of the submillimeter emission
at the obscured side ($\equiv$ $T_{obs}$) can be written as,
\begin{equation}
T_{obs} = T_{re} e^{-\tau} + J(T_{\rm ex}) (1-e^{-\tau}) - J(T_{bg}),
\end{equation}
where
\begin{equation}
J(T)=\frac{\frac{h\nu}{k}}{\exp(\frac{h\nu}{kT})-1}.
\end{equation}
In the above expressions, $h$ is the Planck constant, $k$ is Boltzmann's constant, $\nu$ is the line
frequency, $T_{\rm ex}$ and $\tau$ are the excitation temperature and the optical depth in the foreground cloud,
$T_{bg}$ is the microwave background radiation temperature (= 2.725 K), and
$T_{re}$ is the brightness temperature of the submillimeter emission from the warm cavity wall
at the rear side, respectively. For simplicity, $T_{re}$ is fixed to be the peak brightness temperature
of the halo-like CS emission in L1551 IRS 5 ($\equiv$ 9.2 K; see Fig. \ref{fig:l15comb}). We calculated
$T_{\rm ex}$ and $\tau$ in the foreground cloud with a gas kinetic temperature of 10 K
for a different gas density ($\equiv$ $n_{H_{2}}$) and CS abundance per unit velocity gradient
($\equiv$ $X(CS)/dV/dR$) based on the LVG model, and then derived $T_{obs}$ as a function of the foreground gas density and the
abundance. Details of our LVG calculation are described in Paper I.
In Figure \ref{fig:lvg} we show calculated $T_{obs}$ and $\tau$ as a function of $n_{H_{2}}$
for different $X(CS)/dV/dR$. A horizontal dashed line in the upper panel
shows the 3 $\sigma$ upper limit of the combined ASTE + SMA CS (7--6) data ($\sim$3.6 K).
This figure illustrates that if the foreground cloud is optically thick against the submillimeter CS emission,
the emission from the rear side could be significantly absorbed, down to the sensitivity limit of the
ASTE + SMA CS (7--6) data. The range of the foreground gas density where the line
intensity along the line of sight is lower than the sensitivity limit depends on adopted $X(CS)/dV/dR$.
The typical $X(C^{34}S)/dV/dR$ value in cold dark clouds was estimated to be
$\sim$5.0 $\times$ 10$^{-11}$ km $^{-1}$ s pc from the multi-transitional C$^{34}$S (1--0, 2--1)
observations \citep{tak00}. Then the typical $X(CS) / X(C^{34}S)$ ratio of 22 \citep{chi96}
yields $X(CS)/dV/dR$ = $X(C^{34}S)/dV/dR$ $\times$ 22 $\sim$10$^{-9}$.
With $X(CS)/dV/dR$ = 10$^{-9}$ the density range is $\sim$10$^{5.5-6.3}$ cm$^{-3}$,
which is typical in cloud cores, and the optical depth ranges from $\sim$1 to $\sim$14.

\begin{figure}
\begin{center}
\FigureFile(100mm,100mm){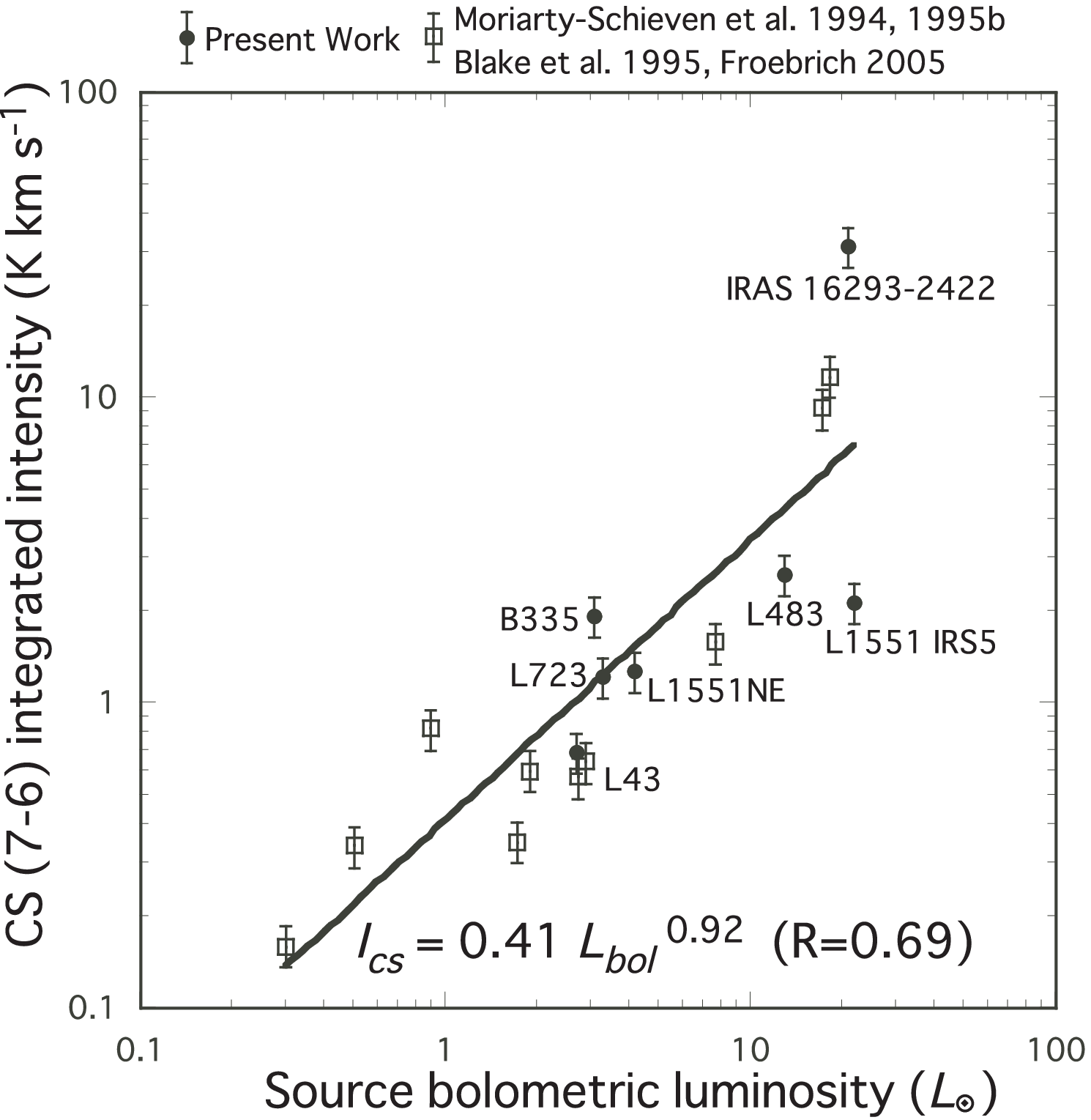}
\end{center}
\caption{Correlation diagram between bolometric luminosities of the protostellar sources and
integrated intensities of the submillimeter CS (7--6) emission. The data points with filled circles
and labels of source names are from Tables \ref{tab:source} and \ref{tab:spectra} in the present paper.
Other data points are from Blake et al. (1995), Moriarty-Schieven et al. (1994, 1995b), and
Froebrich (2005).
A bold line shows the result of the linear fitting to all the data points.}\label{fig:corr}
\end{figure}

Two-dimensional radiative transfer models of protostellar envelopes \citep{spa95,nak03,whi03}
are likely to support our interpretation that the extended submillimeter molecular emissions
at a few thousands AU scale trace walls of the cavity of the envelopes evacuated by the associated outflows.
\citet{spa95} suggested that the narrow $^{12}$CO and $^{13}$CO $J$=6--5 emissions detected
toward many low-mass young stellar objects are produced at the
wall of the cavity evacuated by the bipolar outflow,
heated by the radiation field generated in the inner part of the accretion disk.
\citet{nak03} and \citet{whi03} demonstrated that at the surface of the wall of the cavity opened by the outflow in
the envelope the gas temperature becomes higher than
that predicted from spherically symmetric models, due to the heating by
the protostellar photons.
In fact, the intensities of the submillimeter molecular lines observed with the ASTE 22$\arcsec$ beam
appear to correlate with the central protostellar luminosities, as discussed in $\S$3.1.
To quantitatively show this trend, we compiled published CS (7--6) data \citep{bla95,mo95b}
and protostellar luminosities \citep{mor94,fro05}. In Figure \ref{fig:corr}
we plot integrated intensities of the CS (7--6) emission versus
bolometric luminosities of the protostellar sources.
The data points with filled circles are from Tables \ref{tab:source} and \ref{tab:spectra} in the present paper.
% The CS intensities of other data points are from observations of protostellar sources in Taurus with the CSO 10-m telescope
% by Moriarty-Schieven et al. (1995b), and the bolometric luminosities are from
% Moriarty-Schieven et al. (1994) and Froebrich (2005).
There is a clear linear correlation
between the source bolometric luminosities and the submillimeter CS intensities,
and the least square fitting to the data points yields $I_{CS}$ $\propto$ $L_{bol}^{0.92}$.
The linear correlation between the source luminosities and the intensities of the submillimeter molecular emission
may also support the idea
that the submillimeter molecular emission at a few thousands AU scale
traces the wall of the envelope cavities heated by the protostars.
Detailed physical and chemical models of the envelopes and the 3-dimensional radiative transfer calculations
should help to understand the distributions and the origins of the submillimeter molecular lines in
protostellar envelopes.

% radiative transfer models
% Detailed 3-dimensional radiative transfer models of protostellar envelopes, including the physical and chemical evolution
% of the star-forming envelopes
% and the central protostars, are required to further discuss the correlation
% between the source luminosities and the submillimeter molecular-line intensities, and the distribution
% of the submillimeter molecular lines in protostellar envelopes.

% Bontemps et al. (1996) suggested that the bolometric luminosity is related to the outflow momentum flux, i.e.,
% $F_{CO}$ $\propto$ $L_{bol}$, and hence combing these two relations yields $F_{CO}$ $\propto$ $I_{CS}^{1.8}$.
% The surface of the envelope cavity is likely striped by the associated outflows,
% and the physical properties at the surface are likely related to the outflow activities.
% The apparent correlation between the intensity of the submillimeter molecular emission at
% $\sim$3000 AU scale and the source bolometric luminosity may be due to the correlation
% of the outflow activities and the source luminosities.

\subsection{Different Kinematics between the Millimeter and Submillimeter Molecular Lines in the Protostellar Envelopes}

In Paper I, we suggested that in L483 and B335
the submillimeter molecular lines show opposite velocity gradients to those of the millimeter lines
and the outflows along the outflow axis.
As mentioned in $\S$3.2.1, from our new ASTE data we found that the submillimeter CS emission
in L1551 IRS 5 is also likely to show a similar, opposite velocity gradient.
In this final subsection, we will show how significant these opposite
velocity gradients in the submillimeter molecular line are with our statistical analyses, and discuss the origin of the opposite
velocity gradients.

\begin{figure}
\begin{center}
\FigureFile(150mm,150mm){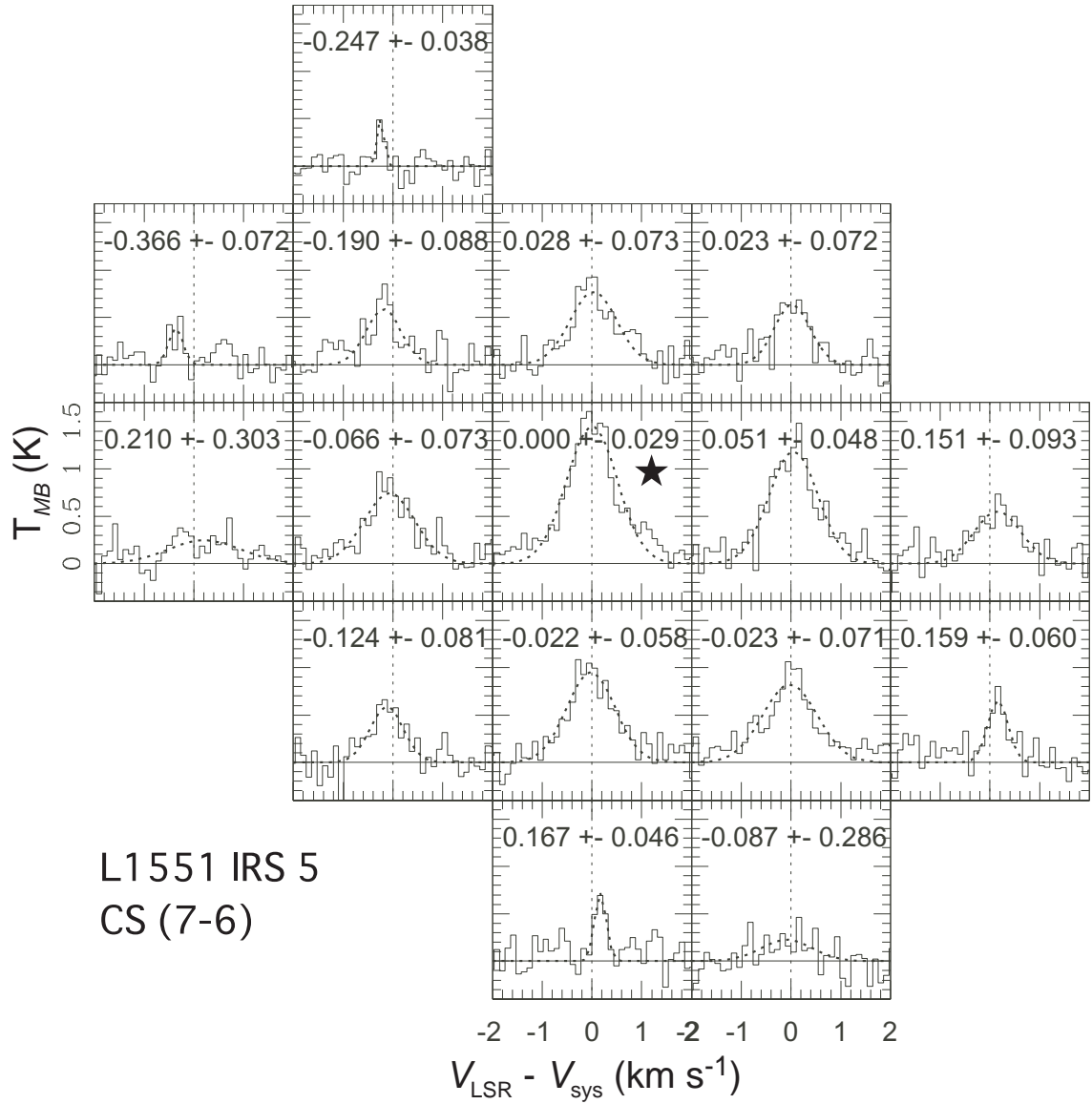}
\end{center}
\caption{Result of the Gaussian fitting to the CS (7--6) spectra in L1551 IRS 5
to derive the distribution of the centroid velocity. Histograms show the observed
line profiles, while dashed curves fitted Gaussians. The horizontal axis shows
the relative velocities with respect to the systemic velocity (vertical dashed lines)
measured toward the protostellar position (star mark). Horizontal solid lines show
the zero intensity level, and the derived relative velocity is shown in each panel.
The grid spacing is 10$\arcsec$, and the upper side is north and the right west.
}\label{fig:l1551csfit}
\end{figure}

\begin{figure}
\begin{center}
\FigureFile(120mm,120mm){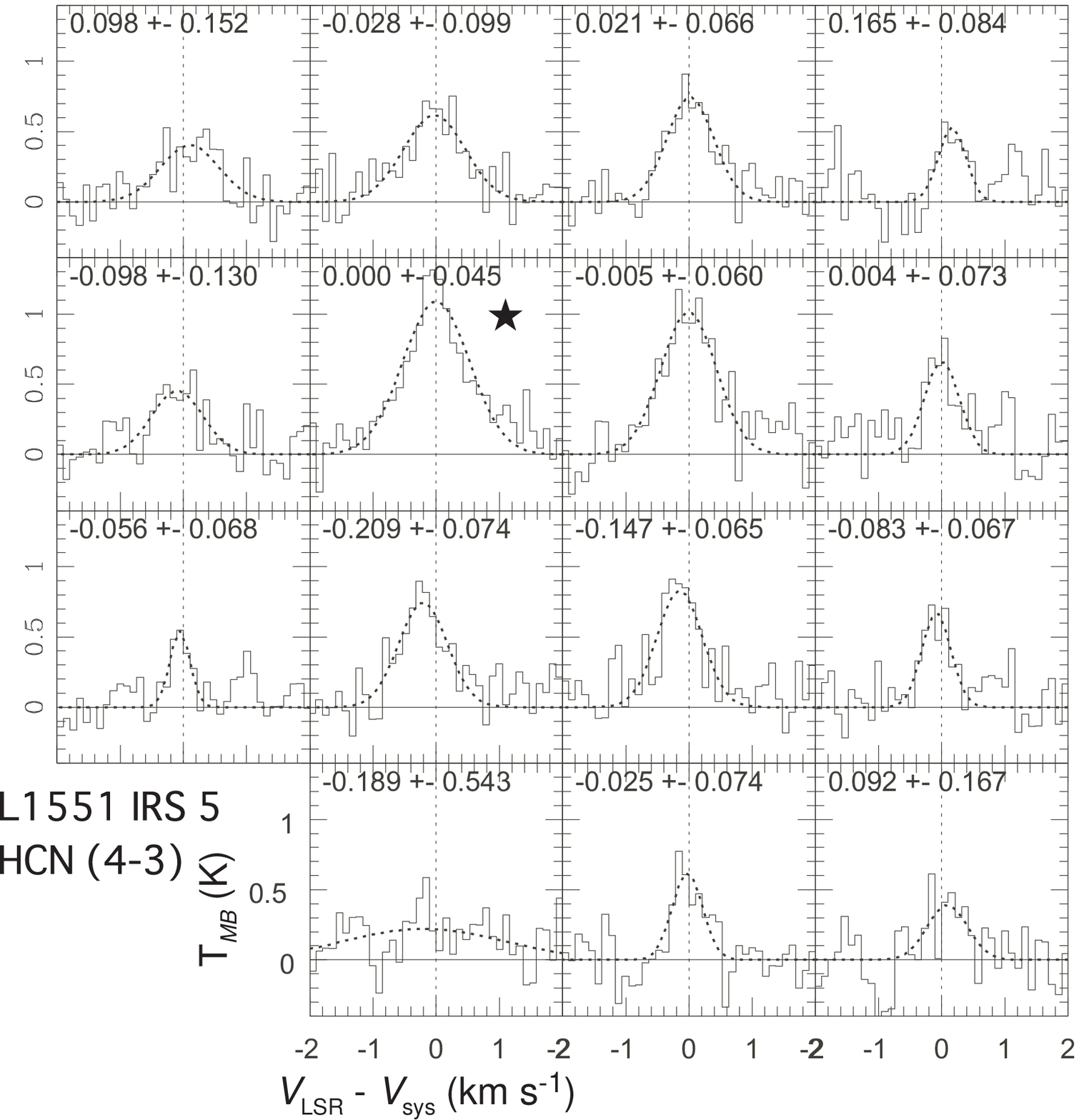}
\end{center}
\caption{Same as Figure \ref{fig:l1551csfit} but for the HCN (4--3) emission.}\label{fig:l1551hcnfit}
\end{figure}

\begin{figure}
\begin{center}
\FigureFile(120mm,120mm){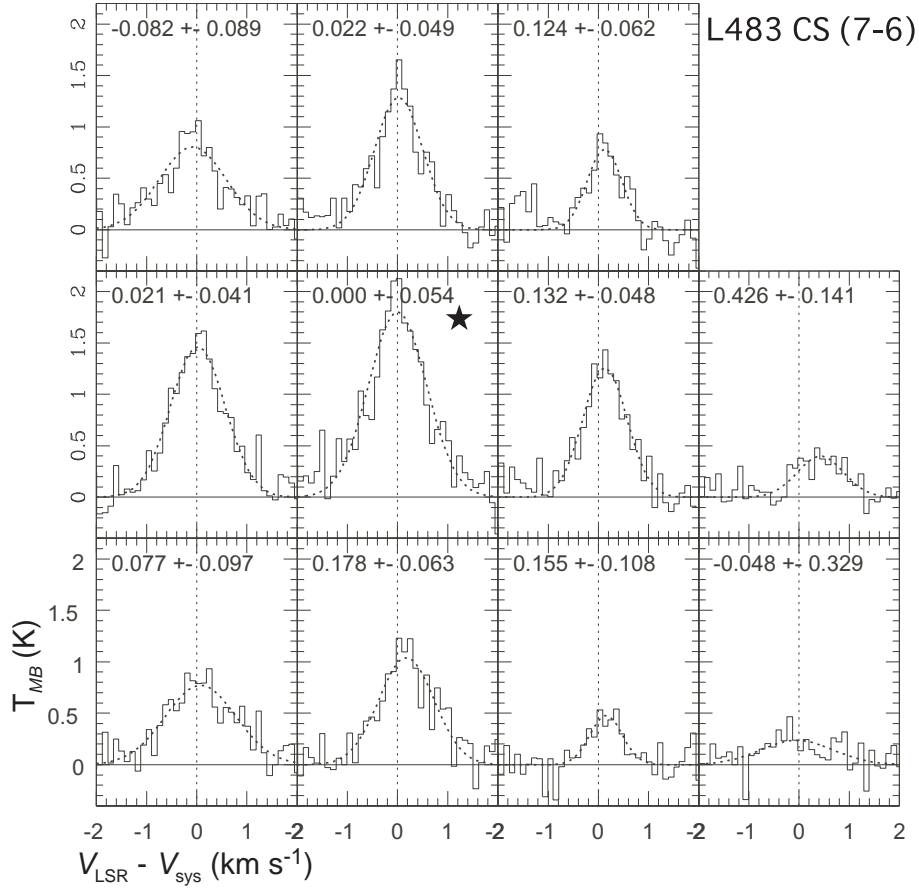}
\end{center}
\caption{Same as Figure \ref{fig:l1551csfit} but for L483.}\label{fig:l483csfit}
\end{figure}

\begin{figure}
\begin{center}
\FigureFile(120mm,120mm){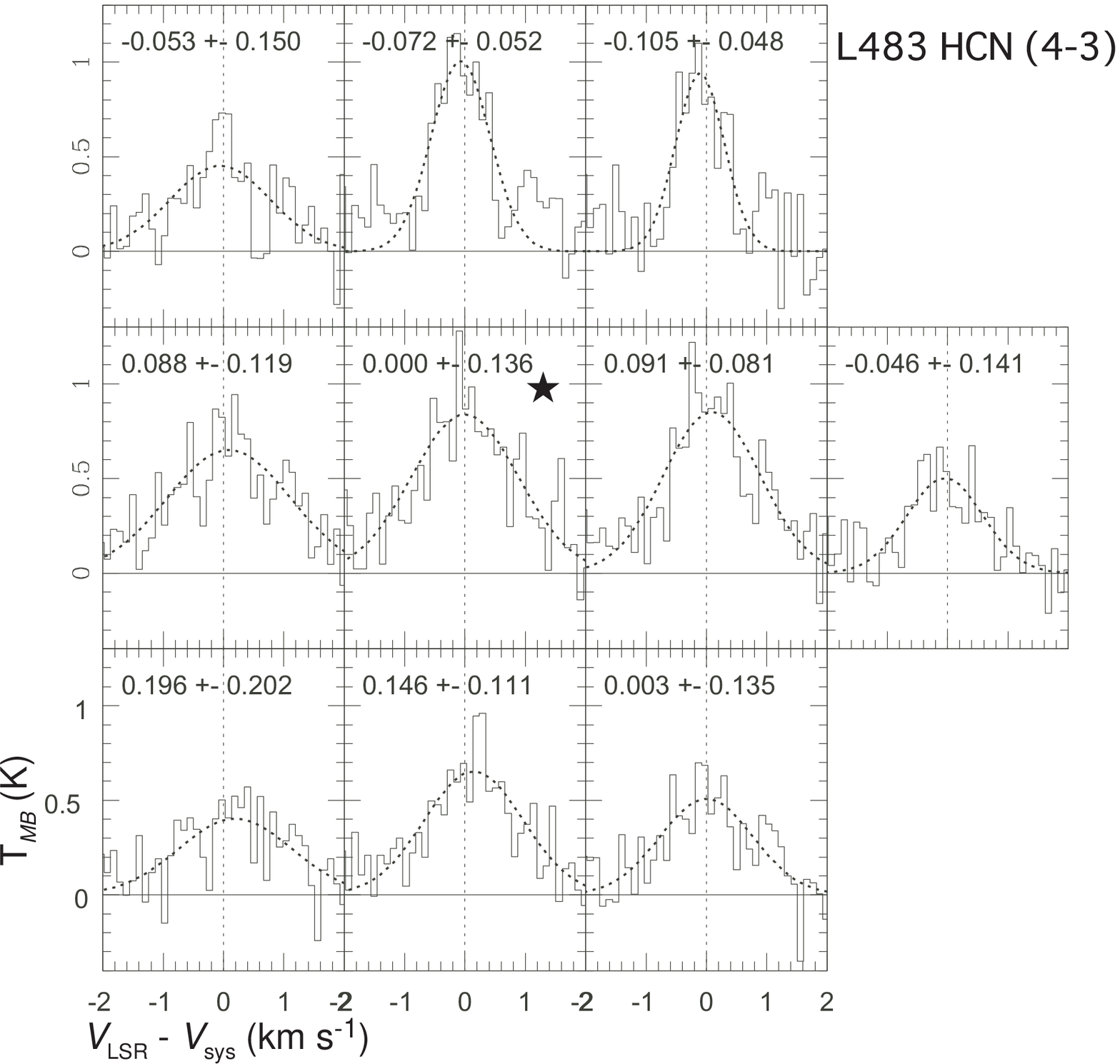}
\end{center}
\caption{Same as Figure \ref{fig:l483csfit} but for the HCN (4--3) emission.}\label{fig:l483hcnfit}
\end{figure}

\begin{figure}
\begin{center}
\FigureFile(100mm,100mm){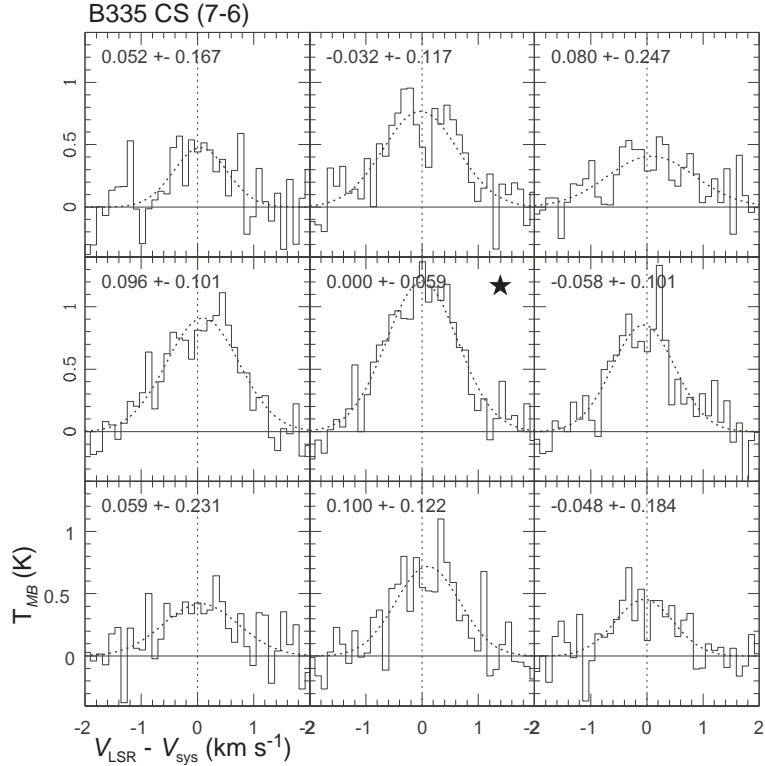}
\end{center}
\caption{Same as Figure \ref{fig:l1551csfit} but for B335.}\label{fig:b335csfit}
\end{figure}

To unambiguously assess the statistical significance of the velocity gradients, we adopted a method
described by Goodman et al. (1993). We first fit a single Gaussian function to each spectrum whose peak intensity
is above 3 $\sigma$, to derive the central $LSR$ velocity of each spectrum ($\equiv$ $v_{LSR}$). The error of the derived central
velocity ($\equiv$ $\sigma_{v_{LSR}}$) can be estimated as
\begin{equation}
\sigma_{v_{LSR}} = 1.15(\frac{\sigma_T}{T})(\delta_v \Delta v)^{\frac{1}{2}},
\end{equation}
where $T$ is the peak of a Gaussian fit to the line profile, $\Delta v$ is the FWHM line width
of a Gaussian fit, $\sigma_T$ is the rms noise in the spectrum, and $\delta_v$ is the ASTE velocity
resolution \citep{lan82}.
Then, we fit a plane function to the estimated central velocities with the errors at the different positions as,
\begin{equation}
v_{LSR} = v_{0} + a\Delta \alpha + b\Delta \delta,
\end{equation}
where ($\Delta \alpha$, $\Delta \delta$) denotes the positions of the spectra with respect to the central protostellar position
along $R.A.$ and $Decl.$, $v_{0}$ is the central velocity of the spectrum toward the central
protostellar position, or the systemic velocity, and $a$ and $b$ are the velocity gradients along
$R.A.$ and $Decl.$, respectively.
In the case of L483 and B335 the outflow axes are along the $R.A.$ direction,
and hence $a$ and $b$ denote the velocity gradients along and across the outflow direction,
respectively.
In the case of L1551 IRS 5 we rotated the vector ($a$, $b$) by -18$^{\circ}$ to estimate
the velocity gradients along and across the outflow axis.
These fittings enable us to estimate the velocity gradients
and the statistical errors along and across the outflow direction
unambiguously. Figures \ref{fig:l1551csfit}, \ref{fig:l1551hcnfit}, \ref{fig:l483csfit},
\ref{fig:l483hcnfit}, and \ref{fig:b335csfit} show the results of the Gaussian fittings to the CS and HCN spectra
in L1551 IRS 5, CS and HCN spectra in L483, and the CS spectra in B335, respectively. Table \ref{tab:plane}
summarizes the results of the plane fittings.

Figure \ref{fig:l1551csfit} shows that the CS spectra in L1551 IRS 5 at the north-eastern side are blueshifted
whereas those at the south-western side redshifted. This sense of the velocity gradient is
opposite to the sense of the infalling envelope and the associated outflow.
The estimated velocity gradient along the outflow direction is
$\sim$(-9.7$\pm$1.7) $\times$ 10$^{-3}$ km s$^{-1}$ arcsec$^{-1}$ (here the negative sign denotes the opposite velocity gradient
to that of the associated outflow and the millimeter molecular lines).
The derived velocity gradient is $\sim$5.9 times the statistical error, and hence
the presence of the velocity gradient in the CS emission opposite to that of the millimeter lines and
the outflow is statistically significant.
In L483, Figure \ref{fig:l483csfit} shows that the CS emission tends to be blueshifted at the eastern side
and redshifted at the western side. The estimated value of the velocity gradient along the outflow direction is
(-7.6$\pm$2.4) $\times$ 10$^{-3}$ km s$^{-1}$ arcsec$^{-1}$, $\sim$3.2 times the statistical error.
No statistical significance of the opposite velocity gradient in the CS (7--6) emission was obtained in B335.

Contrary to the CS (7--6) emission, no significant velocity gradient along any direction was verified
in the HCN (4--3) emission (Table \ref{tab:plane}).
One possible interpretation of the absence of the velocity gradients in the HCN (4--3) emission
is the blending from the hyperfine components. As already mentioned in $\S$3.1.,
two of the six hyperfine components, $F$ = 3--3 and $F$ = 4--4, are shifted by -1.67 km s$^{-1}$
and 1.36 km s$^{-1}$ from the main component ($F$ = 4--3), respectively. The typical line widths measured from the
CS (7--6) line in those protostellar envelopes are 1 - 2 km s$^{-1}$, and hence the 
$F$ = 3--3 and $F$ = 4--4 hyperfine components could totally smear the velocity gradient
in the main $F$ = 4--3 component if the intensity of those satellite lines are comparable
to that of the main line. The much broader HCN
line widths than the CS line widths found in L483 and L723 could also be due to the presence of the hyperfine lines.

From these statistical analyses,
we suggest that at least the CS (7--6) line in L1551 IRS 5 and L483
shows opposite velocity gradients to those of the millimeter lines and the associated outflows.
It is therefore intriguing to discuss the origin of these opposite velocity gradients in the submillimeter line.
As discussed in the last subsection the submillimeter molecular emission likely traces
the warm surface of the cavity opened by the outflow in the envelope,
whereas the millimeter emission the colder, infalling region at the midplane.
% The velocity structures seen in the millimeter emission are interpreted as
% infalling gas motions in the envelopes \citep{sai96,mom98,par00,ye10a}.
Therefore, the submillimeter molecular line should trace a distinct gas motion at the cavity wall
in the envelopes.
A possible interpretation of the observed velocity structures seen in the submillimeter line is
an expanding gas motion which is perpendicular to the outflow axis.
The surface of the cavity wall in the envelope
traced by the submillimeter emission could be
stripped away by some mechanism such as the stellar wind.
As shown in Figure \ref{fig:envconf}, on the side of the blueshifted outflow the
bulk of the expanding gas at the cavity wall must be observed as a redshifted
component and on the side of the redshifted outflow as a blueshifted component.
On the other hand,
the infalling gas component shows the same velocity gradient as that of the outflow.
Hence, this configuration could explain
the opposite velocity gradient observed in the submillimeter line
to that of the outflow or infalling gas traced by the millimeter lines.
In L1551 IRS 5, Pyo et al. (2005) have observed low-velocity ($\sim$100 km s$^{-1}$) [Fe II] winds with a wide opening angle of
$\sim$100$^{\circ}$, as well as collimated high-velocity ($\sim$300 km s$^{-1}$) jets
along the polar axis.
%Comparable results were also obtained for the young T Tauri star DG Tau (\cite{pyo03}).
They have suggested that
such a wind with a wide opening angle will be effective in sweeping up
envelope material from the close vicinity of its driving source. The
observed velocity structure of the submillimeter CS line in L1551 IRS 5 may trace such a
gas motion in the envelope, although the spatial scale in our observations
($\sim$2500 AU) is much larger than that of the [Fe II] observations ($\sim$100 - 400 AU).

On the other hand, previous SMA observations of protostellar envelopes in submillimeter molecular lines
have found compact ($\lesssim$500 AU) components associated with the central protostars, and those components
show mainly rotational gas motion around the protostars \citep{tak04,ta07b,bri07,lom08,jor09,yen11}.
As already mentioned in the last subsection, our combined ASTE + SMA images of the protostellar
envelopes around L1551 IRS 5 and B335 clearly show both the compact components and the
extended components tracing the reflection nebulae. These results suggest that there are
two distinct submillimeter components around protostellar sources; one is a compact ($\lesssim$500 AU),
rotating gas component in the vicinity of the central protostars, and the other
extended ($\sim$2000 AU) feature probably tracing the warm cavity wall of the envelope irradiated by the central
protostars.
% Higher-sensitivity, higher-resolution observations of protostellar envelopes in submillimeter molecular lines with ALMA,
% sampling both extended and fine-scale structures simultaneously, should unveil the origin and mechanism of submillimeter
% molecular lines in low-mass protostellar envelopes.

\begin{table}
\begin{center}
\caption{Results of the plane fitting to the centroid velocities of the submillimeter spectra}\label{tab:plane}
\begin{tabular}{llcccc}
\hline\hline
Line &Source &$a$\footnotemark[$*$] &$b$\footnotemark[$\dagger$] &$v_{0}$\footnotemark[$\ddagger$] &$rms$\footnotemark[$\S$]\\
              &         &$\times$ 10$^{-3}$ (km s$^{-1}$ arcsec$^{-1}$) &$\times$ 10$^{-3}$ (km s$^{-1}$ arcsec$^{-1}$) &(km s$^{-1}$) &(km s$^{-1}$)\\
\hline
CS ($J$=7--6)       &L1551 IRS 5  &-9.7$\pm$1.7 &-3.6$\pm$1.4 &6.51$\pm$0.03 &0.14 \\
                                 &L483               &-7.6$\pm$2.4 &3.6$\pm$2.8 &5.52$\pm$0.05  &0.14 \\
                                 &B335               &-6.5$\pm$5.8  &-3.1$\pm$6.5 &8.18$\pm$0.06  &0.07 \\
HCN ($J$=4--3)    &L1551 IRS 5 &1.2$\pm$1.7   &6.1$\pm$2.1   &6.34$\pm$0.04 &0.09 \\
                                 &L483               &1.4$\pm$3.4  &8.5$\pm$3.4  &5.62$\pm$0.14 &0.06 \\
  \hline
   \multicolumn{6}{@{}l@{}}{\hbox to 0pt{\parbox{180mm}{\footnotesize 
       \par\noindent
       \footnotemark[$*$]Velocity gradient along the outflow direction. The negative sign denotes
       the opposite velocity gradient to that of the associated outflow.
       \par\noindent
       \footnotemark[$\dagger$]Velocity gradient across the outflow direction. The negative sign denotes
       the opposite velocity gradient to that of millimeter molecular lines.
       \par\noindent
       \footnotemark[$\ddagger$]Systemic velocity measured from the one-component Gaussian fitting to the spectrum
       toward the protostellar position.
       \par\noindent
       \footnotemark[$\S$]Rms of the plane fitting to the centroid velocities measured from the Gaussian fitting to the spectra. 
     }\hss}}

\end{tabular}
\end{center}
\end{table}

\section{Summary}

We have conducted mapping observations of L1551 IRS 5, L1551 NE, L723, and L43,
and single-point observations of IRAS 16293-2422
in the submillimeter CS ($J$ = 7--6) and HCN ($J$ = 4--3) lines with ASTE.
We have analyzed the present new ASTE data as well as our previous ASTE data
of L483, B335, L723, and IRAS 16293-2422
(Paper I) in a systematical way, and have obtained the following results.

\begin{itemize}
\item[1]
We detected the CS and HCN lines toward all the protostellar positions
above 4$\sigma$ level, except for the CS and HCN lines at the protostellar position of L1551 NE and
the CS line at the protostellar position of L43.
The CS intensity ranges from $\lesssim$0.5 K (L1551 NE) to $\sim$9.1 K (IRAS 16293-2422),
and the HCN intensity from $\sim$0.4 K (L723) to $\sim$5.8 K (IRAS 16293-2422).
The CS line is stronger than the HCN line toward all the sources except for L43.
There is a linear correlation between the source luminosities and the intensities of
these submillimeter lines ($I_{CS}$ $\propto$ $L_{bol}^{0.92}$).

\item[2]
Our mapping observations of the protostellar envelopes in the CS and HCN emissions
show that the submillimeter emissions often exhibit ``skewed'' distributions
toward the direction of the associated reflection nebulae.
In the combined ASTE + SMA CS (7--6) image of L1551 IRS 5 there appears
an extended ($\sim$2000 AU) component tracing the associated reflection nebula at the southwest,
as well as a compact ($\lesssim$500 AU) component centered on the protostellar position.
The peaks of the CS and HCN emissions in L1551 NE are not located
at the protostellar position but offset by $\sim$1400 AU toward the west, where the associated
reflection nebula resides. The CS emission in L723 is also skewed toward the
direction of the blueshifted outflow.
These results are consistent with our earlier result in L483, where
the CS and HCN emissions are resolved and show elongation ($>$ 2000 AU)
toward the direction of the associated reflection nebula (Paper I).
We suggest that these skewed submillimeter molecular emissions
at a few thousands AU scale trace the warm ($\gtrsim$40 K) walls of the envelope cavities,
excavated by the associated blueshifted outflows and irradiated by the central protostars directly.
On the other hand, the submillimeter emissions at the other side are obscured due to the absorption
from the cold ($\sim$10 K) and dense ($\sim$10$^{6}$ cm$^{-3}$) foreground envelope material.
The detected linear correlation between the protostellar luminosities and the intensities of the
submillimeter molecular lines may also support this interpretation.

\item[3]
From our statistical analyses, we verified that along the outflow directions
the CS (7--6) emission in L1551 IRS 5 and L483 shows opposite velocity
gradients to those of the millimeter molecular emissions and the associated outflows.
The velocity gradients are estimated to be (9.7$\pm$1.7) $\times$ 10$^{-3}$ km s$^{-1}$ arcsec$^{-1}$
in L1551 IRS 5 and (7.6$\pm$2.4) $\times$ 10$^{-3}$ km s$^{-1}$ arcsec$^{-1}$ in L483.
% These results imply that the submillimeter
% molecular emission traces distinct components from the infalling gas in the envelopes
% and the outflows.
One possible interpretation on the origin of the different velocity gradients
is that the submillimeter molecular line traces the dispersing gas motion
at the surface of of the cavity opened by the outflow in the envelope, which is perpendicular to the outflow direction.
The absence of any clear velocity gradient in the HCN (4--3) emission may be due to the
presence of the hyperfine components and the blending of the velocity features.
\end{itemize}

\bigskip

% Acknowledgment should be placed at end of main text.
% (NOT after the Appendix.)

We are grateful to N. Ohashi, P. T. P. Ho, and Masao Saito for their fruitful discussions.
We thank all the ASTE staff for their dedicated support of the telescope
and observatory operations. Observations with ASTE were carried out 
remotely from Japan by using NTT's GEMnet2 and its partner R$\&$E
(Research and Education) networks, which are based on AccessNova 
collaboration of University of Chile, NTT Laboratories, and National Astronomical 
Observatory of Japan. A part of this study was financially
supported by the MEXT Grant-in-Aid for Scientific 
Research on Priority Areas No. 15071202.
S. T. acknowledges the grant from the National Science Council of
Taiwan (NSC 99-2112-M-001-009-MY3) in support of this work.

%%%
% See the manual for the detail.
%%%

\end{document}